\documentclass[twocolumn, superscriptaddress, nofootinbib]{revtex4}

\usepackage{graphicx}
\usepackage{amsmath}
\usepackage{amsfonts}
\usepackage{amsthm}
\usepackage{stmaryrd}
\usepackage{hyperref}

\usepackage{enumitem}
\usepackage{xcolor}

\usepackage{algorithm}
\usepackage{algorithmic}

\makeatletter
\def\algbackskip{\hskip-\ALG@thistlm}
\makeatother

\theoremstyle{plain}
\newtheorem{thm}{Theorem}
\theoremstyle{plain}
\newtheorem{lem}{Lemma}

\theoremstyle{definition}
\newtheorem{defn}{Definition}
\newtheorem{exmp}[thm]{Example}

\theoremstyle{remark}

\newenvironment{proof1}{%
	\proof}{\endproof}

\begin{document}

\title{Parallelized quantum error correction with fracton topological codes}

\author{Benjamin J. Brown}
\affiliation{Centre for Engineered Quantum Systems, School of Physics, University of Sydney, Sydney, New South Wales 2006, Australia}

\author{Dominic J. Williamson}
\affiliation{Department of Physics, Yale University, New Haven, CT 06520-8120, USA}

\begin{abstract}
Fracton topological phases have a large number of materialized symmetries that enforce a rigid structure on their excitations. Remarkably, we find that the symmetries of a quantum error-correcting code based on a fracton phase enable us to design decoding algorithms. Here we propose and implement decoding algorithms for the three-dimensional X-cube model. In our example, decoding is parallelized into a series of two-dimensional matching problems, thus significantly simplifying the most time consuming component of the decoder. We also find that the rigid structure of its point excitations enable us to obtain high threshold error rates. 
Our decoding algorithms bring to light some key ideas that we expect to be useful in the design of decoders for general topological stabilizer codes. Moreover, the notion of parallelization unifies several concepts in quantum error correction. We conclude by discussing the broad applicability of our methods, and we explain the connection between parallelizable codes and other methods of quantum error correction. In particular we propose that the new concept represents a generalization of single-shot error correction.
\end{abstract}

\maketitle

\section{Introduction}

Scalable quantum computation will require us to control the state-space of a system with very high precision, even if some of its physical components experience errors. To this end, we have discovered quantum error-correcting codes; quantum systems with some non-local degrees of freedom that are robust to local noise~\cite{Terhal15, Campbell17}. Among the best quantum error-correcting codes that we know of are those based on topological phases of matter~\cite{Kitaev03, Brown16}. The non-local order parameters of these models naturally give rise to protected quantum states. 
Exploring more general topological phases is a promising route towards the discovery of better codes. In addition to this, it is also important to find new ways to exploit different properties of phases of matter to find more robust procedures for quantum error correction.


Fracton topologically ordered phases~\cite{Vijay15,Vijay16} are remarkable due to the glassy dynamics~\cite{Chamon05, Haah11, Castelnovo11, Bravyi11b,Bravyi13a,PhysRevB.95.155133} of their point-like excitations that are energetically constrained to follow specific trajectories.  Notably, fracton models are structured such that they give rise to a significant number of global materialized symmetries~\cite{Kitaev03, Roberts18, PhysRevB.97.134426}. These are due to relations among the stabilizer generators where the product of a nontrivial subset gives identity. 
Materialized symmetries generally occur when a physical symmetry is gauged~\cite{wegner1971duality, Vijay16, PhysRevB, kubica2018ungauging, you2018symmetric, shirley2018FoliatedFracton}, this connection is explained in detail in Ref.~\cite{Kitaev03}. 
A well-studied example is the {X-cube} model; a three-dimensional model that has a series of two-dimensional materialized symmetries~\cite{PhysRevB.81.184303, Vijay16}.

Here, we show that we can use the materialized symmetries of topological phases to design decoding algorithms. A decoder is essential to the function of a quantum error-correcting code. Its purpose is to take syndrome data and determine the error incident to the code. In this work we propose a decoder for the X-cube model. Remarkably, our example shows that we can decode this model by matching pairs defects of the syndrome on two-dimensional planes of the three-dimensional model in parallel, thus significantly speeding up the process. We attribute this surprising feature to the materialized symmetries of the model. We expect that our method of decoding will be adaptable to any code whose materialized symmetries are known. Given that decoders based on matching typically have high thresholds, we believe our method will enable us to find high-performance decoding algorithms for more general classes of codes. We argue that our approach is particularly well suited to topological codes.

 In what follows, after describing the X-cube model and the implementation of our decoding algorithms, we go on to discuss the prospect of parallelizing other codes, and how the notion of parallelization relates to other modes of decoding. In particular, we propose that single-shot error correction is a special case of parallelized quantum error correction in the fault-tolerant setting.

\section{Model}

\begin{figure}
\includegraphics{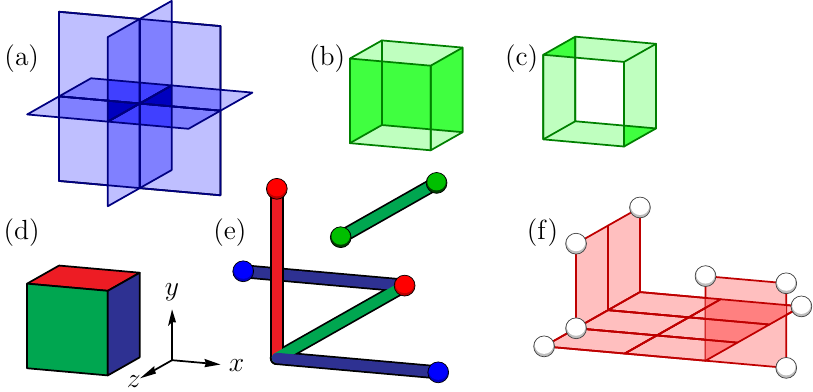}
\caption{The support of stabilizer generators $A_v$, $B^{\mathbf{r}}_c$ and $B^{\mathbf{g}}_c$ are shown in~(a), (b) and~(c), respectively. The faces of the lattice can be three-colored as in~(d). (e)~Lineon excitations which are created by applying Pauli-X errors over dual lines. The color of the excitation is determined by the orientation of the line operator that creates the excitation.  (f)~Fracton excitations of the vertex stabilizers caused by Pauli-Z operators supported on the red faces. \label{Fig:Stabilizers}}
\end{figure}

The X-cube model~\cite{Vijay16} is defined on a cubic lattice with linear size $L$ and periodic boundary conditions. It has a single qubit on each of its faces, and its stabilizer group~\cite{GottesmanThesis} is generated by vertex stabilizers, $A_v$, and cell stabilizers, $B_c^{\mathbf{r}}$ and $B_c^{\mathbf{g}}$, for vertices $v $ and cells $c$ of the lattice. The support of $A_v$, shown in Fig.~\ref{Fig:Stabilizers}(a), includes all the qubits lying on faces that have vertex $v$ on their boundary, i.e., $A_v := \prod_{f,\,  \partial f \ni v} X_f$ with $X_f$ and $Z_f$ the standard Pauli operators acting on $f$. We note that our definition differs from its original description where it is defined on the dual to our cubic lattice~\cite{Vijay16}.

It is helpful to three-color the faces of the lattice to define the cell operators. A face $f$ is colored red, green or blue, denoted $\mathbf{c}(f) = \mathbf{r}, \,\mathbf{g},\,\mathbf{b}$, if it is aligned with the $zx$-, $xy$- or $yz$-plane, respectively, see Fig.~\ref{Fig:Stabilizers}(d). 
A cell operator of color ${\mathbf{c} = \mathbf{r},\,\mathbf{g},\,\mathbf{b}}$ is then defined to have support on all qubits on the boundary of the cell except for the faces of color $\mathbf{c}$, i.e. $B_c^{\mathbf{c}} := \prod_{f \in \partial c,\, \mathbf{c}(f) \not=\mathbf{c}} Z_f $.
  We remark that the stabilizers of a cell form an overcomplete generating set since $B_c^{\mathbf{r}} B_c^{\mathbf{g}}  B_c^{\mathbf{b}}  = 1 $. 
   We show the support of $B_c^{\mathbf{r}}$ and $B_c^{\mathbf{g}}$ in Figs.~\ref{Fig:Stabilizers}(b) and~(c), respectively.\footnote{A commuting stabilizer model is obtained this way on any lattice whose cells have three-colorable faces. We get a phase equivalent to two copies of the three-dimensional toric code if we apply the prescription to a lattice that has four colorable cells~\cite{Bombin07a, Kim11, Brown16a}.} The codespace is the common $+1$ Eigenspace of the elements of the stabilizer group. The mutually anticommuting pairs of logical operators that generate rotations about the codespace, $\overline{X}_j$ and $\overline{Z}_j$ with $ 1 \le j \le 6L-3$  can be chosen to have string-like support~\cite{Vijay16}.

We next look at the structure of the syndrome data of the X-cube model. A syndrome is a list of defects that lie on vertices and cells of the lattice such that $A_v |\psi \rangle  = (-1) |\psi \rangle$, or $B_c^\mathbf{c} |\psi \rangle  = (-1) |\psi \rangle$ for two of the three color labels. A cell defects is assigned a color $\mathbf{c}$ if $B_c^\mathbf{c} = +1$ and the other two stabilizers at cell $c$ are violated. The reader more familiar with condensed-matter physics may think of a defect as an excitation of the Hamiltonian ${H = -\sum_v A_v - \sum_c (B_c^{\mathbf{r}} + B_c^{\mathbf{g}})}$~\cite{Brown16}. Vertex defects, commonly referred to as fractons, are created by Pauli-Z errors, and cell defects, or lineons, by Pauli-X errors. We decode the two types of defect separately.

A pair of cell defects are created by a Pauli-X error incident to a face of a cell. The defects are colored depending on the face the error occurs on. An error on a red, green, or blue face will create a pair of red, green or blue defects, respectively. If two or more errors occur on the faces of the same cell the color of the defect respects a $\mathbb{Z}_2 \times \mathbb{Z}_2$ fusion rule. In other words, similar to the color code~\cite{Bombin06}, the combination of two differently colored defects gives rise to a defect of the remaining distinct color. For instance, the combination of an $\mathbf{r} $ and a $\mathbf{g}$ defect gives a $\mathbf{b}$ defect at the cell where they meet. We can therefore draw strings of errors along edges of the dual lattice, with a color that is determined by their orientation, where the corners must respect the color fusion rules, see Fig.~\ref{Fig:Stabilizers}(e).


Importantly, the X-cube model possesses a materialized $\mathbb{Z}_2$ cell-defect conservation symmetry on planes of the dual lattice that are orthogonal to a coordinate axis. We define the color $\mathbf{c}(\tilde{\Gamma})$ of a dual lattice plane $\tilde{\Gamma}$ in similar fashion to the faces of the lattice. Specifically, the color of any $zx$-, $xy$-, or $yz$-plane is given by $\mathbf{r}, \,\mathbf{g},$ or $\mathbf{b}$, respectively. 
Hence, for any dual lattice plane $\tilde{\Gamma}$ we have a relation $\prod_{c \in \tilde{\Gamma}} B_c^{\mathbf{c}(\tilde{\Gamma})} = 1$ by taking the product of all the cell operators of color $\mathbf{c}(\tilde{\Gamma})$. 
For example, the product of all the cell operators of color $\mathbf{r}$ lying on a plane of constant $y$ will return identity. 

It follows that the total number of defects of two different colors is conserved modulo 2 on a plane of the third distinct color. 
For instance, given that both red and green defects require that $B_c^\mathbf{b} = -1$, on blue planes $\tilde{\Gamma}$, where $\prod_{c \in \tilde{\Gamma}} B_c^\mathbf{b} = 1$, we have that the parity of the number of red and green defects must be equal.

Finally, we briefly describe the vertex defects. Vertex defects are created on the vertices at the corners of the faces affected by a Pauli-Z error. Further, vertex defects respect a $\mathbb{Z}_2$ fusion rule. In Fig.~\ref{Fig:Stabilizers}(f) we show a Pauli-Z operator with the vertex defects it creates marked by white vertices. Similar to the cell defects, the vertex defects respect a materialized $\mathbb{Z}_2$ symmetry on each of the $xy$-, $yz$-, and $zx$-lattice planes, $\Gamma$. Specifically, we have the relation $\prod_{v \in \Gamma}  A_v = 1$ for any such $\Gamma$.

\section{Decoding with symmetries}

A decoding algorithm takes a syndrome that consists of a collection of defects caused by some Pauli error $E$, and then attempts to return a correction operator $C$ that will restore the code to its initial state.
We say that a subset, or cluster, of defects can be neutralized if they can be created by a Pauli error supported on qubits that are nearby the cluster.
We look for small clusters of defects that can be neutralized with a low-weight Pauli operator. With this strategy, in the limit that the error rate is small, it is highly likely that $CE$ is a member of the stabilizer group.  Our strategy to find neutralizable sets is to pairwise group nearby defects within specifically chosen subregions of the lattice that reflect the materialized symmetries of the model. The intuition behind this approach is that defects are created in clusters such that all materialized symmetries are respected. 
Conversely, in what follows we find that neutralizable clusters are obtained by ensuring that every defect of the cluster is paired to another multiple times; once for each materialized symmetry of which the defect is a member. The performance of the decoder is determined by its ability to conduct well-informed pairing during each of the pairing subroutines.

We expect that with some imagination most strategies for decoding can be adapted to perform the pairing subroutines~\cite{Harrington, Duclos-Cianci10, Wootton12, Bravyi13a, Anwar14, Torlai17}. Of particular interest is the decoder due to Delfosse and Nickerson~\cite{Delfosse17} that is readily adapted to the problem at hand to complete each matching subroutine in almost linear time. Moreover this decoder, and others similar~\cite{Harrington, Bravyi13a, Anwar14}, should permit a generalization of our work to qudit stabilizer codes. Here we make pairings using the Kolmogorov implementation~\cite{Kolmogorov09} of the minimum-weight perfect matching algorithm due to Edmonds~\cite{Edmonds65}. The algorithm takes a graph with weighted edges and returns a subset of the edges of the original graph such that each vertex has exactly one incident edge and the sum of the weights of the edges that are returned is minimal. We use this algorithm by assigning each defect a vertex of the graph, and adding edges to the graph that are weighted to approximate the logarithm of the likelihood that a Pauli-error caused the pair of defects that are connected by the edge~\cite{Dennis02, Stace10, Fowler12b, Hutter14a, Nickerson17, Criger18}.

Most interestingly, all the decoders we present use matching subroutines on the defects lying on $L\times L$ planes of the system rather than its full volume. Parallelization here significantly speeds up our algorithm. The subroutines could be performed in almost $L^2$ time~\cite{Delfosse17}, thus enabling us to decode the X-cube model in sublinear time.

\begin{figure}
\includegraphics{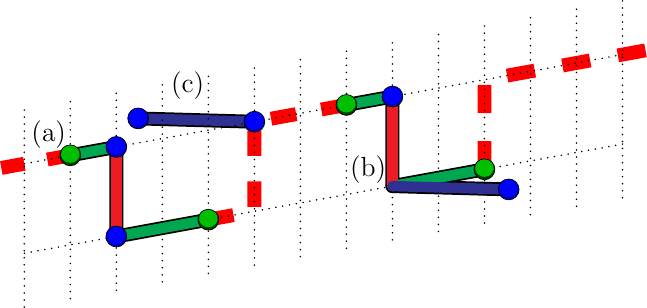}
\caption{A Pauli-X error. On the plane marked by the dotted grid we match all the green and red excitations. We use the blue excitations to help find the correct matching. (a)~shows an error where two green defects are caused a string like error with blue defects at its corners. Nearby errors can lead to confusion. At~(b) an error moves a blue defect away from the plane, and at~(c) an additional error introduces a blue defect that could also mislead the decoder into finding an incorrect matching along the red dashed line. \label{Fig:MatchingStrategy}}
\end{figure}

We first propose a decoder that neutralizes cell defects. Our decoder matches all of the defects of each plane that are constrained by a planar $\mathbb{Z}_2$ symmetry. 
For instance we match the red and green defects on each plane of constant $x$. Likewise we match the green and blue (red and blue) defects on every plane of constant $y$ ($z$). 
The connected components of the defects, where connections are given by edges  returned from all $3L$ different matching calculations, are neutral clusters. We show this explicitly in Appendix~\ref{App:CellDecoder}. Here we focus on a single plane to better understand the performance of the decoder.

In Fig.~\ref{Fig:MatchingStrategy} we show a plane upon which we perform matching. Here the red and green defects are paired. The number of blue defects is not conserved on this plane. If a red and a green defect meet at a common cell they are expended to create a blue defect, see Fig.~\ref{Fig:MatchingStrategy}(a) where a local error creates two blue defects. Single blue defects can also be created on or moved from the plane by applying errors to qubits on faces that are parallel to the plane of interest. See Figs.~\ref{Fig:MatchingStrategy}(b) and~(c) where a single blue defect is created on the plane, and a blue defect is moved away from the plane, respectively.

Our proposed decoder that neutralizes vertex defects gives a threshold of $\sim 9.4\%$ for an independent and identically distributed noise model. Notably, our decoder makes use of the rigid structure of the defects in our fracton code. Specifically, we use the locations of the blue defects as a `breadcrumb trail' to assign more accurate weights to the edges in the red and green defect matching. This gives the decoder for the fracton model an advantage as the locations of the blue defects reduce the degeneracy of the lowest weight string errors that is typically present in two-dimensional models~\cite{Stace10, Duclos-Cianci10, Criger18, Beverland18}. 

To illustrate the potential advantages that are to be gleaned from the rigid excitations of the X-cube model we show an error that can be corrected by a decoder that uses information about the blue defects in Fig.~\ref{Fig:MatchingStrategy}. We first remark that a standard decoder that assigns weights according to the separation of the defects will not be able to deal with this error. This is because the sum of the edge weights along the incorrect path, shown by the red dashed line, is equal to the sum of the edge weights along the path that would yield a correction that successfully restores the system to its initial state. 

In contrast, we find the decoder that accounts for the blue defects will succeed on this particular example. The path of the successful correction  matches the red and green defects via three blue defects, leaving only one corner without a blue defect. The edges that connect the defects along the incorrect path must include two corners with no blue defect. As such, the minimum-weight matching decoder that accounts for blue defects during the pairing will deal with this error with certainty. 
To exploit this additional information offered by the blue defects we penalize the pairing of red and green defects around corners than do not support a blue defect. We describe the implementation of this decoder in Appendix~\ref{App:CellDecoder}.

\begin{figure}
\includegraphics{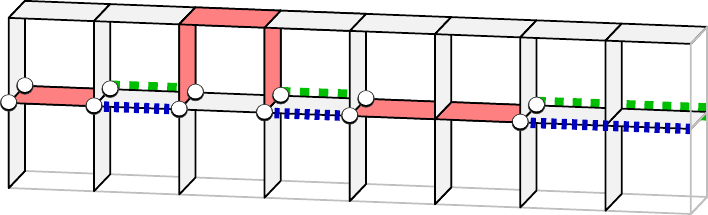}
\caption{Pauli-Z errors on a plane of the lattice. By comparison with the toric code we see this error cannot be corrected. \label{Fig:LogicalErrors}}
\end{figure}

We decode the vertex defects by pairing each of them with another defect that shares its $x$ coordinate, a defect that shares its $y$ coordinate, and a defect that shares its $z$. We show that a cluster that respects this pairing can be neutralized in Appendix~\ref{App:VertexDecoder}. Therefore, conducting pairing subroutines on the $xy$-, $yz$- and $zx$- planes connects groups of defects to give neutralizable subsets. Performing matching on these planes, where we suppose the likelihood of an error creating a pair of defects is proportional to their separation, gives a threshold of $\sim 3.8\%$.

We also argue that the threshold should not exceed ${\sim 10\%}$. We consider the qubits on the plane shown in Fig.~\ref{Fig:LogicalErrors}. If we project these qubits into the page, we effectively have the edges of the square lattice toric code where the syndrome is duplicated twice; once at the front of the image and once at the back.
If errors only occur on these qubits, then we expect that the decoder will make incorrect pairings with high probability at the threshold error rate of the toric code, $\sim10\%$~\cite{Dennis02}. Given that our argument only considers errors on one plane of qubits we should not expect that this bound is tight.

The defects of the X-cube model have a lot of structure due to a subextensive number of global materialized symmetries. We can improve the decoder by communicating outcomes of the previous iterations of the matching subroutine to find more accurate weightings for later rounds of pairing. We detail the implementation of our iterative decoder in Appendix~\ref{App:VertexDecoder}. Our improved decoder presents a threshold of $\sim 4.4 \%$. We suggest that more refined schemes of belief propagation should be applicable to design very high threshold decoders with fracton codes. However, given the upper bound we have obtained for this particular model, optimization of the X-cube model is unlikely to be of practical value. 

It is interesting to compare our results to the threshold for the point defects of the three-dimensional toric code, $\sim 3.3\%$~\cite{Kubica17}. Unlike the point-defects of the X-cube model, these defects are constrained by only a single global materialized symmetry. As such, its syndrome is considerably less structured. Following similar reasoning, it is unsurprising that our thresholds fall short of the high threshold the three-dimensional toric code presents against bit-flip errors, $\sim 23.5\%$~\cite{Kubica17}. Indeed, the stabilizer defects that arise due to bit-flips are highly constrained by an extensive number of local materialized symmetries. Nevertheless, we believe that the encouraging thresholds we have obtained here, together with the potential for parallelization, motivates further study of fracton topological codes in the context of active quantum error correction. Additional exploration may lead to the discovery of codes with very high thresholds.

\section{Other examples of parallelized quantum error correction}

Parallelized quantum error correction is a concept we have introduced that unifies a number of developments in decoding algorithms. We give examples of how symmetries have been used to design other decoders in Appendix~\ref{App:Relations}. Moreover, we have proposed several new tools that enable us to decode highly symmetric models with fracton topological order. Below we discuss parallelizable codes in the broader context of quantum error correction, and motivate new directions of study.

We first compare parallelizable codes with single-shot error correction~\cite{Bombin15a, Brown16a, Campbell18, fawzi2018constant}. Typically, in the fault-tolerant setting where measurements are unreliable, we collect syndrome data over a long period and decode once we have a sufficiently large history~\cite{Dennis02}. Models with single-shot error correction can be decoded within planes of constant time without an extensive syndrome history. Notably, as we explain in Appendix~\ref{App:TimeCorrelated}, the resilience of single-shot codes to time-correlated errors~\cite{Bombin16} can be understood from the perspective of parallelized quantum error correction. In contrast, the X-cube decoder we present can be parallelized over spatially distinct planes. In this sense, we might regard single-shot error correction as a special case of parallelization. This analogy becomes clearer still in the fault-tolerant setting. Indeed, one can conceive of decoding schemes for the four-dimensional spacetime history of the X-cube model where matching is performed on three-dimensional materialized spacetime symmetries. Foliation might be used to compare these error-correction protocols evenhandedly~\cite{Bolt16, Brown18}.

Beyond these models, self-correcting memories~\cite{Brown16} are both single-shot~\cite{Bombin15a} and fully parallelizable due to their local materialized symmetries. As such, these memories are readily decoded using cellular automata~\cite{Dennis02, Pastawski11, Breuckmann17, Kubica18}. Like single-shot codes then~\cite{Bombin15a}, we might regard parallelizable codes as an intermediate class of codes between self-correcting models and two-dimensional codes that do not readily permit parallelization.

\section{Summary and conclusions}

A key idea in our work is the use of defect pairing on subsets of defects that respect materialized symmetries of the model. Although a similar method of decoding has been applied to a number of low-dimensional stabilizer codes elsewhere in the literature, to the best of our knowledge this generic method has not been explicitly identified before. As discussed in  Appendix~\ref{App:Relations}, we expect our approach is broadly applicable to topological stabilizer codes. Further study may even reveal our method applies to general stabilizer codes where the symmetries of the model are known.

The decoding framework we have presented refines the clustering decoder~\cite{Bravyi13a} where clusters are grown around defects until the cluster respects all of the symmetries of the model. In contrast, we deal with each symmetry separately. Further, our refinement permits the use of minimum-weight perfect matching, which tends to yield high thresholds. As such, we hope our work inspires the discovery of new high-performance decoders for other codes. Fracton topological codes offer a rich playground to begin such an exploration. Our decoder should carry over directly to other foliated fracton models~\cite{shirley2017fracton, shirley2018FoliatedFracton, shirley2018Fractional, Slagle2018}. Also of interest are the type-II fracton models which admit no string operators~\cite{Haah11, Vijay16}. This class includes the cubic code~\cite{Haah11,Bravyi11b,Bravyi13a} which, notably, has logical operators with better distance scaling than the surface code. Due to its fractal structure, which leads to partial self-correction, we might expect this model to give rise to advantageous properties~\cite{Nixon20}.

\begin{acknowledgements}
We Acknowledge S. Bartlett for supportive and inspiring conversations, N. Delfosse for helpful discussions on the complexity of decoding algorithms and to A. Kubica for offering a simplified explanation of the color code decoders by projection. The authors acknowledge the facilities of the Sydney Informatics Hub at the University of Sydney and, in particular, access to the high performance computing facility Artemis. BJB is supported by the University of Sydney Fellowship Programme and the Australian Research Council via the Centre of Excellence in Engineered Quantum Systems(EQUS) project number CE170100009. 
\end{acknowledgements}

\appendix

\setcounter{secnumdepth}{1} 

\section{The cell defect decoder}
\label{App:CellDecoder}

Here we describe in detail the decoder that corrects Pauli-X errors which give rise to cell defects. We first argue that collections of defects that are connected via the proposed pairing subroutines will give rise to neutralizable subsets. We then describe our implementation of the decoder that achieves the pairing, and how we use the minimum-weight perfect matching algorithm to account for corner defects. We describe our decoder with pseudocode in Appendix~\ref{App:Pseudo}.

We perform many minimum-weight matching subroutines. We first establish some terminology where we divide the subroutines into one of three subsets, either the red, green or blue matching, where the color depends on the orientation of the plane on which the matching subroutine is performed. Minimum-weight perfect matching is performed on each plane between all the defects that have a different color to the plane, as these defects respect a parity conservation symmetry. Specifically, each defect of color red and green will be paired with either a red or a green defect with the same $x$ coordinate, and defects of color red and blue (green and blue) will be paired with another defect of color red or blue (green or blue) that has the same $z$ ($y$) coordinate. We say that the collection of edges that connect red and green defects on any plane of constant $x$ are the result of the blue matching. Likewise, the collection of edges that connect red and blue (green and blue) defects on any plane of constant $z$ ($y$) are the result of the green (red) matching.

We now define a correctable cluster of  cell defects, and then go on to argue that the correctable cluster is neutralizable.
\begin{defn}[Correctable cluster of cell defects] 
A cluster of cell defects is correctable if for every defect included in the subset the subset also includes the defect it is paired with on each of the three differently colored matchings; the red, the green, and the blue matching.
\end{defn}
We remark that a defect may be paired with the same defect in two different matchings. For instance, a green defect may be displaced from another green defect along the $z$ axis and share a common $x$ and $y$ coordinate. In which case, these two defects may be paired in both the red and blue matching. This gives a correctable cluster.

Given that the union of the edges returned from all of the minimum-weight perfect matching subroutines divides the syndrome into correctable clusters, it remains to show that we can find a Pauli operator that will neutralize a correctable cluster of defects. We therefore prove the following theorem.

\begin{figure}
\includegraphics{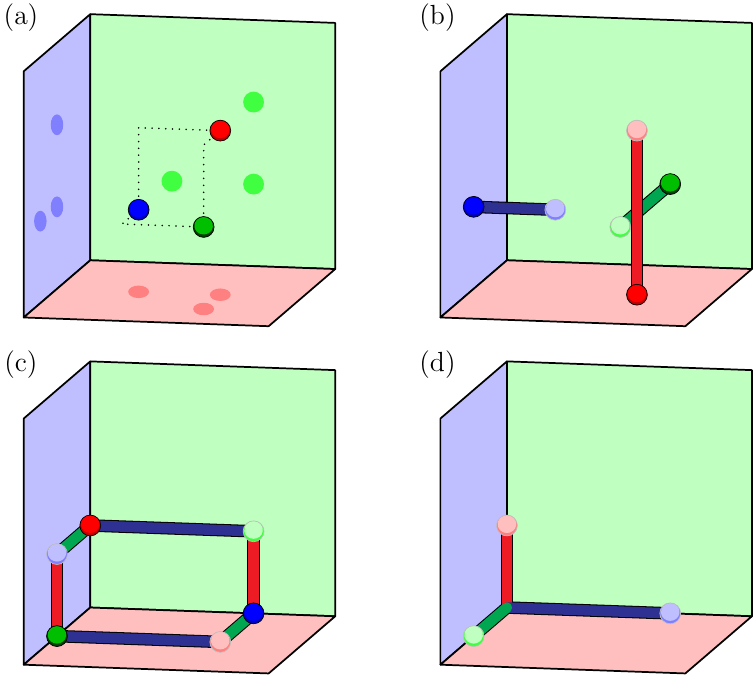}
\caption{\label {Fig:LineonCorrection} The process of finding a correction operator for a correctable collection of cell defects. (a)~An odd parity of each of the differently colored subsets of defects is shown in the bulk of the lattice. (b)~Each of the defects are moved freely onto a plane. (c)~The defects are then moved onto a one-dimensional line where the planes meet. Finally, (d)~the defects are moved along their respective one dimensional lines to neutralize at a common point. }
\end{figure}

\begin{thm} \label{Thm:CorrectingLineons}
We can efficiently find a Pauli operator that neutralizes a correctable cluster of defects.
\end{thm}
The constructive proof of the theorem is sketched in Fig.~\ref{Fig:LineonCorrection}. The proof relies on the fact that a correctable cluster has an equal parity of each of the three differently colored cell defects. We therefore prove the following lemma before proceeding with the proof of Theorem~\ref{Thm:CorrectingLineons}.

\begin{lem} 
A matched correctable cluster of defects has an equal number of defects of each color modulo 2. 
\end{lem}

\begin{figure}
\includegraphics{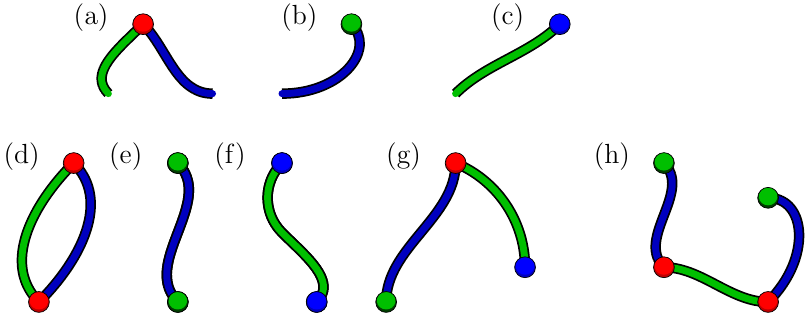}
\caption{\label{Fig:Edges} Showing that the parity of the number of defects of the same color is equal. Each red defect has one incident edge from the green matching and one from the blue matching~(a), and the green and blue defects have only one incident edge from the blue and green matching, see~(b) and~(c), respectively. These form one-dimensional chains where an even number of red defects can form a one dimensional cycle~(d), two greens or two blues can be connected directly~(e) and~(f), a string of an odd number of connected red defects terminates at two differently colored defects~(g) or an even number of red defects can terminate at two defects of the same color~(h). The number of differently colored defects in each of the possible combinations that are shown explicitly between~(d) and~(h) are of equal parity.}
\end{figure}

\begin{proof}
To prove the lemma we consider the edges of the blue and green matching between the defects of a correctable cluster. Each red defect is paired to two other defects; one defect in the green matching and another defect through the blue matching. Similarly, each green (blue) defect is paired to only one other defect through the blue (green) matching. As such, the connectivity of each vertex with respect to their color is shown precisely as in Figs.~\ref{Fig:Edges}(a),~(b) and~(c), where edges are colored to indicate the color of the matching the edge came from.

We exhaustively consider the connected components of the defects of a correctable cluster that are connected via edges of the green and the blue matching. We first remark that the connected components are necessarily one dimensional, where a string of red defects may form a cycle of an even number of defects, or, possibly, a chain terminates with either green or blue defects. Our exhaustive examination will show that the numbers of defects of each color in any connected component have equal parity.

One eventuality is that a connected component gives rise to a one-dimensional cycle of reds. The alternating edge type guarantees the number of reds in the cycle is even, such that the parity of all the defects in this chain is equal, see Fig.~\ref{Fig:Edges}(d). It is also possible for a pair of greens or blues to be connected directly through their respective pairings, see Figs.~\ref{Fig:Edges}(e) and~(f). The number of defects of every color in each of these three cases is even.

We next consider connected components of the subset where the red defects form longer chains that are not cyclic. We find that the edges that are taken from differently oriented planes guarantees that the parity of differently colored vertices still remains equal. For instance, in Fig.~\ref{Fig:Edges}(g) the alternation of edges is such that a one-dimensional string of an odd number of red defects terminates at a green defect at one end and a blue defect at the other. In this case we have that the number of defects of each color in this connected component of the subset will be odd.

Alternatively, as we see in Fig.~\ref{Fig:Edges}(h), an even number of red defects in a terminating chain guarantees the terminal defects of the chain are of the same color, thus leading to an even number of defects of each colors in this connected component of the subset. This completes all of the cases.

Given that the representative connected components shown in Figs.~\ref{Fig:Edges}(d)-(h) have equal parity of each color of defect, it follows that the union of any of these connected components must also have equal parity.  This concludes the proof of the lemma.
\end{proof}

We remark that the argument above explains why the color code decoder due to Delfosse~\cite{Delfosse14, KubicaThesis} will give rise to connected components that can be locally corrected.

Using the fact above, we argue that we can find a Pauli operator that will neutralize a correctable cluster of defects. To do so, we continually reduce the dimensionality of the problem. For example, see see Fig.~\ref{Fig:LineonCorrection} where we show a configuration with an odd parity of red, green and blue defects such that each defect shares a plane with another defect and consequently the defect conservation symmetry is respected on all planes. Important to the proof is ensuring that the parity of the number of cell defects of each color remains equal throughout.

\begin{proof1}
To find a correction, we first specify three planes. We move defects onto these planes to reduce the dimensionality of the problem. We choose one plane of constant $x$, $y$ and $z$ as shown in blue, red and green, respectively, in Fig.~\ref{Fig:LineonCorrection}. We first move all of the defects onto the plane of their respective color, see Fig.~\ref{Fig:LineonCorrection}(b). This is achieved with a Pauli operator that will not create any additional cell defects. Moreover, the parity of the number of each subset of colored defects is conserved. 

It may be that two defects of the same color are moved to a common point of a given plane. These two defects are neutralized. It remains to consider how the other defects connected to the two neutralized defects are paired. We find that we can consistently pair the two defects adjacent to the two neutralized defects.

For concreteness, we name the defects $\delta_j$ with $ 1 \le j \le 4$ such that $\delta_j $ is paired to $\delta_{j+1}$ for $j \le 3$. It may be that $\delta_1$ and $\delta_4$ are paired to other defects that we are not interested in. Defects $\delta_2 $ and $\delta_3$ share the same color, $\mathbf{u}$, and meet at a common point after the first move. Let's say that $\delta_2$ and $\delta_3$ are paired through the $\mathbf{v}$-colored matching with $\mathbf{v} \not= \mathbf{u}$. Now, $\delta_1$ is necessarily paired to $\delta_2$ through the $\mathbf{w}$-colored matching with $\mathbf{w} \not= \mathbf{u},\mathbf{v}$. Likewise, $\delta_3 $ is paired to $\delta_4$ via the same $\mathbf{w}$-colored matching. Given that $\delta_2$ and $\delta_3$ met at a common point after the initial move, it follows that $\delta_1$ and $\delta_4$ can be consistently paired via the $\mathbf{w}$-colored matching, and the remainder of the proof will continue as follows. Moreover, given that a pair of defects of the same color are neutralized, the parity of the number of defects of that color is not changed.

We also remark that two defects of different colors may meet at a point where two planes intersect. Or three differently colored defects meet at the point where all three defects intersect. All of these eventualities respect the color parity of each of the differently colored defects. In the former case, two defects of different colors $\mathbf{u}$ and $\mathbf{v}$ fuse to give a defect of color $\mathbf{w}$. This maintains the color parity of the defects.  In the latter case one defect of each color is neutralized. This also maintains consistency among the differently colored subsets.

After the move what remains on each of the planes is an equal parity of each of the differently colored subsets of defects. The next step is to move all the defects onto the lines where the colored planes intersect. See Fig.~\ref{Fig:LineonCorrection}(c). Pairs of differently colored defects that met at these lines of intersection can skip the following step. Otherwise, we require an `L-shaped' Pauli operator, that divides a defect of color $\mathbf{u} =\mathbf{r},\, \mathbf{g},\, \mathbf{b}$ into two defects, colored $\mathbf{v},\mathbf{w} \not= \mathbf{u}$, and move the two charges towards the plane of their respective color, $\mathbf{v}$, $\mathbf{w}$, respectively. The result of applying the L-shaped correction will leave an equal parity of the differently colored defects on each of the lines where the differently colored planes intersect. From here the remaining defects on these lines can be neutralized at the intersection point of all three of the planes as in Fig.~\ref{Fig:LineonCorrection}(d).

We finally argue that the defects that lie on each of these three lines must be of equal parity. Without loss of generality we consider the line that collects defects of color $\mathbf{u}$ where the planes of color $\mathbf{v}$ and $\mathbf{w}$ intersect. We first remark that the parity of defects of color $\mathbf{v}$ and $\mathbf{w}$ on each of the planes of their respectively color are equal. Moreover, every defect of color $\mathbf{v}$ or $\mathbf{w}$ is paired with one other defect of color $\mathbf{v}$ or $\mathbf{w}$ through the $\mathbf{u}$-colored matching.

We recall that pairs of defects that are matched via the $\mathbf{u}$ matching will meet at the same point along the line of intersection between the $\mathbf{v}$ and $\mathbf{w}$ colored plane. If two defects are paired by the same color, they meet at the line, and neutralize along this line. This action preserves the color symmetry of the subset. Altenatively, a pair of defects of color $\mathbf{v}$ and $\mathbf{w}$ combine to give a single defect of color $\mathbf{u}$. In this case, the number of defects of color $\mathbf{v} $ and $\mathbf{w}$ both decrease by one, but we increase the number of defects of color $\mathbf{u}$, thus maintaining the equal parity of defects of a given color. Given that this operation occurs simultaneously on each line of intersection, this completes the argument for neutrality.
\end{proof1}

\subsection{Implementation}

\begin{figure}
\includegraphics{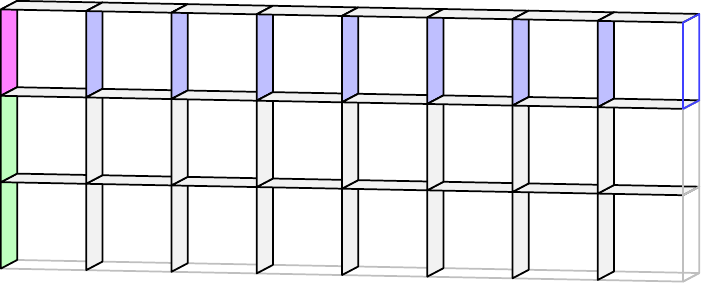}
\caption{The support of the one dimensional logical operators of the X-cube model. The logical Pauli-X (Pauli-Z) operators are supported on the blue (green) qubits, and both logical operators intersect at the pink face. \label{Fig:LogicalOperators}}
\end{figure}

The decoder requires that each defect is paired with another defect on each plane such that all the charge conservation symmetry is conserved. We achieve this with minimum-weight perfect matching~\cite{Edmonds65, Kolmogorov09}. The minimum-weight perfect matching takes as input a list of vertices, that are connected via a list of weighted edges. The algorithm returns a list of edges such that each vertex has one and only one incident edge, and the sum of the weights of the edges that are returned is minimal.

We propose a decoder that accounts for the information provided by the defects that are not conserved on a given plane. We consider the $\mathbf{u}$-colored plane where defects of color $\mathbf{v}$ and $\mathbf{w}$ are conserved. Each of these defects are represented by a vertex in the minimum-weight perfect matching algorithm. Further, we add two vertices for each of the defects of color $\mathbf{u} \not= \mathbf{v},\, \mathbf{w}$ on the plane. We assign weights to edges between all of the vertices proportional to their separation according to their Manhattan distance, except where we increase the weight by one if two defects are separated via a corner. Now, pairs of defects of color either $\mathbf{v}$ or $\mathbf{w} $ can be connected via edges that pass through defects of color $\mathbf{u}$. 
If two such defects are not aligned parallel to a lattice axis the weight of their matching is penalized by one for each corner of the path that does not pass through a defect of color $\mathbf{u}$. Defects that are aligned parallel to a lattice axis can be paired without the corner penalty.

\begin{figure}
\includegraphics[width=\columnwidth]{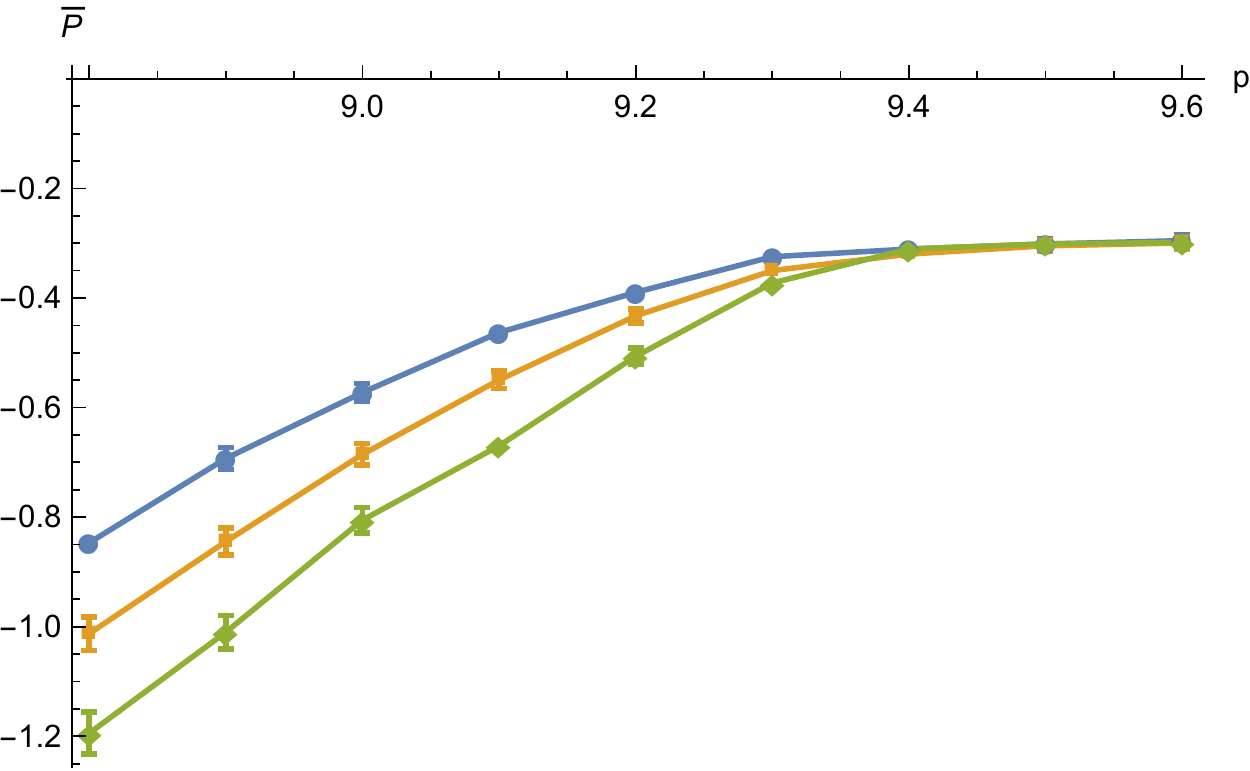}
\caption{Logical failure rate using the decoder to correct cell defects where errors are independent and identically distributed bit-flips at rate $p$. We find a threshold $\sim 9.4\% $ at the crossing point where we compare system sizes $L = 72, \,84,\,96,\,108$. \label{Fig:BitFlipThreshold}}
\end{figure}

We find the logical failure rate by comparing the support of the correction operator our decoder returns to one logical operator as shown in Fig.~\ref{Fig:LogicalOperators}. This decoder shows a threshold of $\sim 9.4\%$, see Fig.~\ref{Fig:BitFlipThreshold}.

\section{Decoding Pauli-Z errors}
\label{App:VertexDecoder}

We next describe the decoder that finds a low-weight Pauli operator to neutralize vertex defects. We briefly summarize how correctable clusters can be found via parallelized matching subroutines, and we explain how a correction operator can be obtained. We then describe the implementation of our decoder and, in particular, explain how we reiterate the matching subroutines to improve our priors on each matching procedure.

We seek clusters of defects such that we can apply a Pauli operator to each cluster that will neutralize all of its defects.  Vertex defects must respect the planar materialized symmetries of the X-cube model, and so too must a neutralizable cluster of vertex defects. To this end,  each defect included within a neutralizable cluster must be paired with precisely one other defect that shares its $x$ coordinate, one that shares its $y$ coordinate, and one that shares its $z$ coordinate.

We achieve the pairing by performing minimum-weight perfect matching subroutines over two-dimensional planes of the three-dimensional model. It is helpful to define the matching subroutines. We call the collection of matching subroutines that pair defects with a common $x$ coordinate the $x$-matching and, likewise, the collection of matching subroutines that pair defects with a common $y$ ($z$) coordinate the $y$-matching ($z$-matching). Every defect of a neutralizable cluster is paired with as many as three other defects via the combination of these matching procedures. It may be that a pair of defects are matched twice through two different matching subroutines if one defect is displaced from the other along a direction parallel to a lattice axis.  This eventuality is consistent with the following discussion. It also follows that every defect must be paired with at least two distinct other defects.

\begin{figure}
\includegraphics{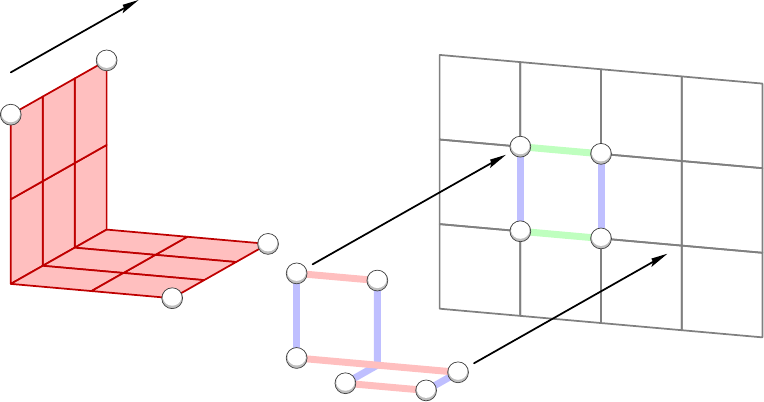}
\caption{\label{Fig:PlaneonCorrection} Correctable clusters of vertex defects. (left)~We can move defects paired on a plane of constant $z$ onto other planes of constant $z$ with the operator shown. (right)~The projection of 
defects that were paired over planes of constant $z$ are shown on a single plane. Given that each defect is paired to another defect on the $xz$ and $yz$ plane, the projected edges mark a boundary with projected defects on its corners. The defects on this plane can be corrected by Pauli-Z operators acting on the faces enclosed by the boundary.}
\end{figure}

We next describe how a correction operator is obtained for each neutralizable cluster. Similar to the case with the cell-defect decoder, we can simplify the problem considerably by projecting the defects of a neutralizable cluster onto a two-dimensional plane. Without loss of generality we project all the defects of a neutralizable cluster onto a plane of constant $z$. This projection is achieved by moving pairs of defects with the same $z$ coordinate using a Pauli operator similar to that shown in the left of Fig.~\ref{Fig:PlaneonCorrection}. Given that every defect in a neutralizable cluster is paired with one other with the same $z$ coordinate, we can necessarily make this projection by the definition of a neutralizable cluster.

It remains to consider the defects that lie on the plane after the projection, see Fig.~\ref{Fig:PlaneonCorrection}(right). We recall that every defect is paired with one other defect with a common $x$ and a common $y$ coordinate by the definition of a neutralizable cluster. This implies that, after the projection, and supposing that the defects are not neutralized at the plane, that each defect is aligned with two other defects; one horizontally and one vertically. In fact, we can consider the edges that pair these defects as a one-dimensional boundary on the plane. Applying a Pauli-Z operator to each of the qubits that lie on the interior of this boundary will neutralize all of the remaining defects.

\subsection{Implementation}

We finally describe our use of minimum-weight perfect matching to pair defects such that we can find neutralizable clusters. We first describe a simplified version where the pairings are all achieved na\"ively. We then describe a generalization of the decoder where we repeat minimum-weight perfect matching several times such that the outcome of the last matching is used to improve the results of the next. We summarize the decoder in terms of pseudocode in Appendix~\ref{App:Pseudo}

\begin{figure}
\includegraphics[width=\columnwidth]{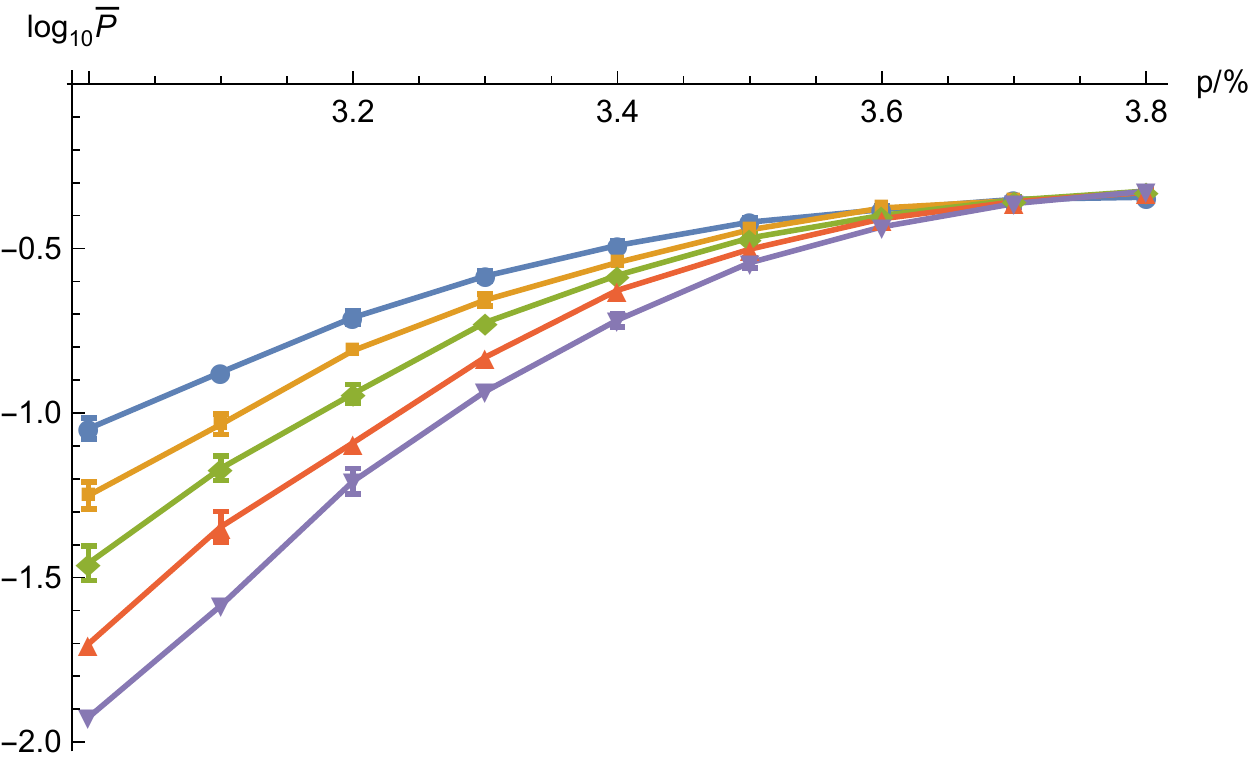}
\caption{Threshold error rate for the vertex excitations where we perform matching over planes that satisfy a $\mathbb{Z}_2$ symmetry, and we assign weights to edges according to the separation of two vertices by their Manhattan distance. We find a threshold $\sim 3.7\% $ using system sizes $L = 36,\, 42,\, 48,\, 54,\, 60$.\label{Fig:PlaneonIter1}}
\end{figure}

A neutralizable cluster as defined above requires that every vertex defect is paired with two or three other vertex defects; one must share its $x$ coordinate, one must share its $y$ coordinate, and one must share its $z$ coordinate. We achieve the pairing with minimum-weight perfect matching. We perform a minimum-weight perfect matching subroutine on each plane of constant $x$, $y$ and $z$ where we input a graph such that each vertex defect on the given plane is assigned a vertex, and we weight each edge that connects two vertices of the complete graph according to the Manhattan distance that separates the two vertices it connects. The vertices are paired according to the output of the edges of each matching subroutine. We obtain a threshold of $\sim 3.7\%$ using this decoder where we compare the correction operator our decoder returns to the logical operator shown in Fig.~\ref{Fig:LogicalOperators}. We collected data using $\sim 10^4$ samples with system sizes $L = 36,\,42,\,48,\,54 $ and $60$ to obtain this result. See Fig.~\ref{Fig:PlaneonIter1}.

It is important to consider how we use minimum-weight perfect matching to find an error that was the most likely to have given rise to a configuration of defects. Ideally, we assign weights to edges such that they are proportional to the logarithm of the probability that an error caused those two defects. For instance, in the case of the surface code, weights are typically assigned according to the separation of the defects, as the weight of the error must increase with their separation if we suppose an independent and identically distributed noise model~\cite{Dennis02}. However, this is a suboptimal solution, as we find the most likely error that caused the syndrome, and an optimal decoder will return a representative correction operator for the class of homologically equivalent errors that most likely caused the syndrome~\cite{Wang03, Duclos-Cianci10}. To overcome this, some references have considered heuristic modifications to the weighting function of edges to account for degeneracy~\cite{Stace10, Criger18, Beverland18} or correlated features~\cite{Fowler12b, Hutter14a, Nickerson17} of the error model.

A local Pauli-Z error gives rise to at least four vertex defects. To decode this correctly, it would be favorable to use minimum-weight hypergraph matching, where we minimize the sub of the weights to hyper edges that connect multiple vertices. In the absence of an efficient hypergraph-matching algorithm, we turn to belief propagation methods to supplement minimum-weight graph matching. We perform minimum-weight perfect matching several times, where we use the output of the last matching to improve the next. Specifically, we have a prior matching where each vertex defect is paired to another via the $x$, $y$ and $z$ matching. We use this prior matching to better estimate the weights for the edges in the subsequent matching. 
The priors are initialized by matching with weights given by the Manhattan distance between pairs of vertices. 
We use the output of the final matching to determine the correction.

\begin{figure}
\includegraphics{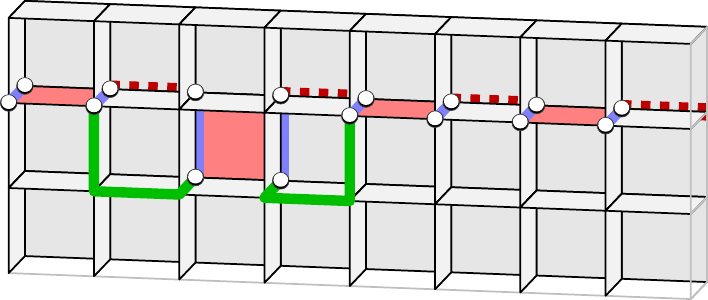}
\caption{\label{Fig:DifficultError} An error configuration marked by red faces that cannot be decoded by pairing defects using minimum-weight perfect matching where edges are weighted according to their separation. The defects may be paired incorrectly via the red dashed lines. We improve the decoder by choosing weights such that we also account for the separation between defects that have paired with the defects of the edge we are trying to assign a weight to. We look at the defects that have been paired via the $x$-matching that are connected to the defects of interest via the blue edges. Due to the difference in orientation of one of the errors, if we assign weights that account for the separation of the defects of interest, and the separation of their other partners, a high weight will be assigned to the incorrect edges. We mark the large separation between the paired defects by green lines.}
\end{figure}

To motivate our choice of weighting function we consider the error shown in Fig.~\ref{Fig:DifficultError}. In the figure, we mark the faces that have experienced errors by coloring them red. The resulting defects could be connected incorrectly with high probability in either the $y$- or the $z$-matching via the red-dotted edges if we choose weights by their separation only. This will lead to a logical failure. However, we note that one of the errors that gives rise to the extensive line of defects is oriented differently from the others. As such, we expect that assigning weights that also account for the locations of other nearby defects may enable us to correct such an error. We communicate the results of different matching subroutines to produce better estimates for edge weights.

\begin{figure}
\includegraphics[width=\columnwidth]{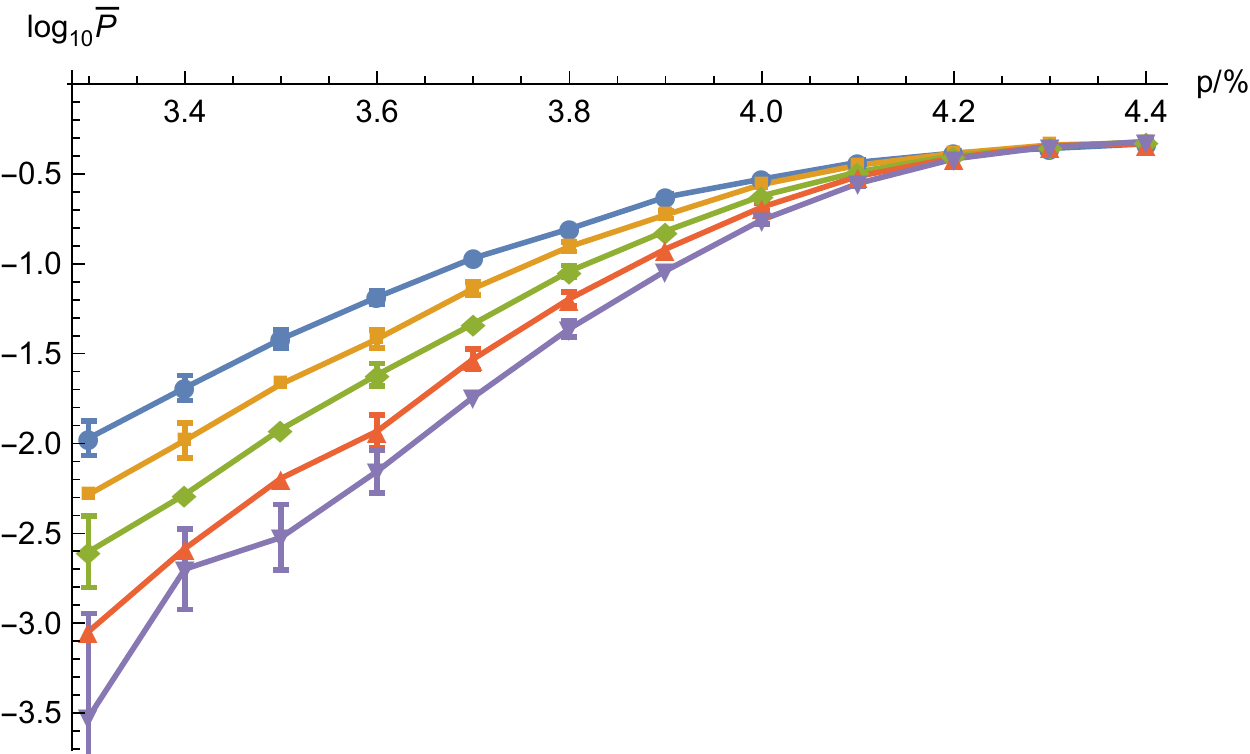}
\caption{Threshold error rate using the vertex defect decoder after five iterations of belief propagation. We find a threshold $\sim 4.3\%$ using system sizes $L = 36,\,42,\,48,\,54,\,60$. We use $\sim 10^4$ samples to find each data point. \label{Fig:Iter5}}
\end{figure}

Once the priors are initialized, we assign weights to edges that connect two vertices using the separation between the other defects a pair of defects are paired to, as well as their own separation. For instance, without loss of generality, we calculate the weight of an edge separating two defects for the $x$-matching. We label these two defects $\delta_1$ and $\delta_2$. The defects are paired to one or two other defects via the priors from the $y$- and $z$-matching. We label the defects paired with $\delta_j$ as $\delta_j^y$ and $\delta_j^z$ respectively for $j = 1,\,2$, and we denote by $\text{sep}(\delta, \delta')$ the Manhattan distance separating defects $\delta$ and $\delta'$. Supposing $\delta_1^y \not= \delta_2$, which also implies $\delta_2^y \not= \delta_1$ and $\delta_1^z \not= \delta_2$, we choose the weight for the edge $e$ connecting $\delta_1$ and $\delta_2$ such that
\begin{equation}
W(e) \sim \text{sep}(\delta_1, \delta_2) + \min [\text{sep}(\delta_1^y, \delta_2^y), \text{sep}(\delta_1^z,  \delta_2^z) ].
\end{equation}
We do not consider the separation between paired defects if they are the same defects as those we are trying to determine the weight between. In the case that, say, $\delta_1^y = \delta_2$, we simply take ${W(e) \sim \text{sep}(\delta_1, \delta_2) + , \text{sep}(\delta_1^z,  \delta_2^z) }$. Likewise, if  $\delta_1^z = \delta_2$, we simply take ${W(e) \sim \text{sep}(\delta_1, \delta_2) + , \text{sep}(\delta_1^y,  \delta_2^y) }$. If $\delta_1^y = \delta_2$, it is necessarily true that $\delta_1^z \not= \delta_2$ and vice versa. As such we always find a defect that is paired that is distinct from the defects of the edge we are assigning a weight to.

We find that the decoder rapidly converges to a higher threshold as we increase the repetitions of the belief propagation step. We find a threshold of $\sim 4.3\%$ if we reiterate five times. This value was collected using $\sim 10^4$ Monte Carlo samples for system sizes $L = 36,\,42,\,48,\,54,$ and $60$. The data is shown in Fig.~\ref{Fig:Iter5}.

\section{Pseudocode}
\label{App:Pseudo}

In this appendix, we summarize the implementation of the decoder we have presented in terms of pseudo code. Both decoders we have discussed, and the decoder we present in the following section follow the overarching procedure presented in Algorithm~\ref{Alg:Overview}.

\begin{algorithm}
\caption{$\mathtt{decode}$ \label{Alg:Overview}}

\begin{algorithmic}
\STATE{\hspace{-0.165in} {\bf Input:} $\textsf{syndrome}$}\\
\STATE{\it // {$\textsf{syndrome}$} is a configuration of defects.}
\STATE{{\hspace{-0.165in} \bf Output:} $\textsf{correction}$} \\
\STATE{\it // {$\textsf{correction}$ is a Pauli correction operator. }}
\STATE

\STATE{\it // Part I - Find neutral clusters.} 
\FORALL{$\Gamma$}
\STATE{$\textsf{edges}(\Gamma) \leftarrow \mathtt{MWPM}(\textsf{syndrome} \cap \Gamma)$}
\ENDFOR
\STATE $G \leftarrow \left( \textsf{syndrome},\, \bigcup_\Gamma \textsf{edges}(\Gamma) \right)$
\STATE {\bf find} $\textsf{clusters}_j(G)$
\STATE{\it // Variables $\Gamma$ denote planar symmetries of the code.}
\STATE{\it // Subroutine $\mathtt{MWPM}$ is minimum-weight perfect matching.}
\STATE{\it // Variables $\textsf{clusters}_j(G)$ are disjoint networks of graph $G$.}

\STATE

\STATE{\it //  Part II - Find a correction for neutral clusters.}

\STATE $\textsf{correction} \leftarrow 1$

\FORALL{$\textsf{clusters}_j(G)$} 
\STATE {$\textsf{correction}_j \leftarrow \mathtt{neutralize}(\textsf{clusters}_j(G)) $}
\STATE $\textsf{correction} \leftarrow \textsf{correction}\times \textsf{correction}_j$
\ENDFOR
\STATE{\it // Subroutine $\mathtt{neutralize}$ takes $\textsf{clusters}_j(G)$ and returns a correction that locally neutralizes the defects of the cluster.}
\RETURN $\textsf{correction}$
\end{algorithmic}
\end{algorithm}

\begin{algorithm}
\caption{$\mathtt{neutralize}$ (lineon) \label{Alg:Lineon}}
\begin{algorithmic}
\STATE{\hspace{-0.165in} {\bf Input:} $\textsf{cluster}$} \\
\STATE{{\hspace{-0.165in} \bf Output:} $\textsf{correction}  \leftarrow 1$} \\
\STATE
\STATE{\it // Part I - Move defects to their respective planes.} 
\FORALL{ $ \textsf{v} \in \textsf{cluster} $}
\STATE $ \textsf{initial pos.} \leftarrow \mathtt{position}(\textsf{v})$
\STATE
\IF{ $\mathtt{color}(\textsf{v}) = \mathbf{r}$}
\STATE $ \textsf{final pos.} \leftarrow ( \mathtt{x} (\textsf{v}) , M^\mathbf{r}, \mathtt{z} (\textsf{v}) ) $
\STATE $ \textsf{correction}  \leftarrow \textsf{correction}\times \mathtt{move}^\mathbf{r}(\textsf{initial pos.} \rightarrow \textsf{final pos.}) $
\STATE
 \ELSIF{$\mathtt{color}(\textsf{v}) = \mathbf{g}$}
\STATE $ \textsf{final pos.} \leftarrow ( \mathtt{x} (\textsf{v}) , \mathtt{y} (\textsf{v}), M^\mathbf{g} ) $
\STATE $ \textsf{correction}  \leftarrow \textsf{correction}\times \mathtt{move}^\mathbf{g}(\textsf{initial pos.} \rightarrow \textsf{final pos.}) $
\STATE
 \ELSIF{$\mathtt{color}(\textsf{v}) = \mathbf{b}$}
\STATE $ \textsf{final pos.} \leftarrow ( M^\mathbf{b},  \mathtt{y} (\textsf{v}) , \mathtt{z} (\textsf{v}) ) $
\STATE $ \textsf{correction}  \leftarrow \textsf{correction}\times \mathtt{move}^\mathbf{b}(\textsf{initial pos.} \rightarrow \textsf{final pos.}) $
\ENDIF
\ENDFOR

\STATE

\STATE{\it //  Part II - Move defects to planar intersections.}
\FORALL{ $ \textsf{v} \in \textsf{cluster} $}
\STATE $ \textsf{initial pos.} \leftarrow \mathtt{position}(\textsf{v})$
\STATE
\IF{ $\mathtt{color}(\textsf{v}) = \mathbf{r}$}
\STATE $ \textsf{final pos. a} \leftarrow ( \mathtt{x} (\textsf{v}) , M^\mathbf{r}, M^\mathbf{g} ) $
\STATE $ \textsf{final pos. b} \leftarrow  ( M^\mathbf{b} , M^\mathbf{r}, \mathtt{z} (\textsf{v}) )$

\STATE $ \textsf{correction}  \leftarrow \textsf{correction}\times \mathtt{move}^\mathbf{g}(\textsf{initial pos.} \rightarrow \textsf{final pos. a}) $
\STATE $ \textsf{correction}  \leftarrow \textsf{correction}\times \mathtt{move}^\mathbf{b}(\textsf{initial pos.} \rightarrow \textsf{final pos. b}) $
\STATE

 \ELSIF{$\mathtt{color}(\textsf{v}) = \mathbf{g}$}
\STATE $ \textsf{final pos. a} \leftarrow ( \mathtt{x} (\textsf{v})  , M^\mathbf{r}, M^\mathbf{g}) $
\STATE $ \textsf{final pos. b} \leftarrow  ( M^\mathbf{b}, \mathtt{y} (\textsf{v})  , M^\mathbf{g})$

\STATE $ \textsf{correction}  \leftarrow \textsf{correction}\times \mathtt{move}^\mathbf{r}(\textsf{initial pos.} \rightarrow \textsf{final pos. a}) $
\STATE $ \textsf{correction}  \leftarrow \textsf{correction}\times \mathtt{move}^\mathbf{b}(\textsf{initial pos.} \rightarrow \textsf{final pos. b}) $
\STATE

 \ELSIF{$\mathtt{color}(\textsf{v}) = \mathbf{b}$}
\STATE $ \textsf{final pos. a} \leftarrow ( M^\mathbf{b} ,M^\mathbf{r}, \mathtt{z} (\textsf{v}) ) $
\STATE $ \textsf{final pos. b} \leftarrow  ( M^\mathbf{b} , \mathtt{y} (\textsf{v}) , M^\mathbf{g})$

\STATE $ \textsf{correction}  \leftarrow \textsf{correction}\times \mathtt{move}^\mathbf{r}(\textsf{initial pos.} \rightarrow \textsf{final pos. a}) $
\STATE $ \textsf{correction}  \leftarrow \textsf{correction}\times \mathtt{move}^\mathbf{g}(\textsf{initial pos.} \rightarrow \textsf{final pos. b}) $

\ENDIF
\ENDFOR

 \STATE

\STATE{\it //  Part III - Move remaining defects to the origin.}
\FORALL{ $ \textsf{v} \in \textsf{cluster} $}
\STATE $ \textsf{initial pos.} \leftarrow \mathtt{position}(\textsf{v})$
\STATE $ \textsf{final pos.} \leftarrow ( M^{\mathbf{b}}, M^\mathbf{r}, M^\mathbf{g} ) $
\STATE

\IF{ $\mathtt{color}(\textsf{v}) = \mathbf{r}$}

\STATE $ \textsf{correction}  \leftarrow \textsf{correction}\times \mathtt{move}^\mathbf{r}(\textsf{initial pos.} \rightarrow \textsf{final pos.}) $

 \ELSIF{$\mathtt{color}(\textsf{v}) = \mathbf{g}$}
\STATE $ \textsf{correction}  \leftarrow \textsf{correction}\times \mathtt{move}^\mathbf{g}(\textsf{initial pos.} \rightarrow \textsf{final pos.}) $

 \ELSIF{$\mathtt{color}(\textsf{v}) = \mathbf{b}$}
\STATE $ \textsf{correction}  \leftarrow \textsf{correction}\times \mathtt{move}^\mathbf{b}(\textsf{initial pos.} \rightarrow \textsf{final pos.}) $

\ENDIF
\ENDFOR

\STATE
\RETURN $\textsf{correction}$
\end{algorithmic}
\end{algorithm}

\begin{algorithm}
\caption{$\mathtt{neutralize}$ (planeon) \label{Alg:Planeon}}
\begin{algorithmic}
\STATE{\hspace{-0.165in} {\bf Input:} $\textsf{cluster}$} \\
\STATE{{\hspace{-0.165in} \bf Output:} $\textsf{correction}  \leftarrow 1$} \\
\STATE
\STATE{\it // Part I - Move defects to a common plane.} 
\FORALL{ $ \textsf{e} \in \textsf{edges}^z $}
\STATE$ \textsf{v\textsubscript{1}} \leftarrow \mathtt{vertex\textunderscore 1}(\textsf{e})$
\STATE $\textsf{v\textsubscript{2}} \leftarrow \mathtt{vertex \textunderscore 2}(\textsf{e})$

\STATE $ \textsf{initial pos.} \leftarrow \{\mathtt{position}(\textsf{v\textsubscript{1}}), \mathtt{position}(\textsf{v\textsubscript{2}}) \}$
\STATE $ \textsf{final pos.} \leftarrow \{ (\mathtt{x}(\textsf{v\textsubscript{1}}), \mathtt{y}(\textsf{v\textsubscript{1}}),M), (\mathtt{x}(\textsf{v\textsubscript{2}}), \mathtt{y}(\textsf{v\textsubscript{2}}),M)  \}$
\STATE $ \textsf{correction}  \leftarrow \textsf{correction}\times \mathtt{move}(\textsf{initial pos.} \rightarrow  \textsf{final pos.}) $
\ENDFOR
\STATE

\STATE{\it // Part II - Find correction boundary.} 
\STATE $\textsf{boundary} \leftarrow 0$
\STATE

 \FORALL{    $\textsf{e}  \not\in \textsf{edges}^z$ }
\STATE$ \textsf{v\textsubscript{1}} \leftarrow \mathtt{vertex\textunderscore 1}(\textsf{e})$
\STATE $\textsf{v\textsubscript{2}} \leftarrow \mathtt{vertex \textunderscore 2}(\textsf{e})$

\STATE $ \textsf{pos. a} \leftarrow (\mathtt{x}(\textsf{v\textsubscript{1}}), \mathtt{y}(\textsf{v\textsubscript{1}}),M) $
\STATE $ \textsf{pos. b} \leftarrow (\mathtt{x}(\textsf{v\textsubscript{2}}), \mathtt{y}(\textsf{v\textsubscript{2}}),M) $
\STATE $\textsf{line segment} \leftarrow \mathtt{draw\textunderscore line}(\textsf{pos. a} \rightarrow \textsf{pos. b} ) $
\STATE $\textsf{boundary} \leftarrow \textsf{boundary} + \textsf{line segment} $
\ENDFOR
\STATE
\STATE{\it // Part III - Correct the interior.} 
\STATE $ \textsf{interior} \leftarrow \mathtt{get\textunderscore interior}(\textsf{boundary})$

\FORALL{$\textsf{qubits} \in \textsf{interior}$}
\STATE $\textsf{correction} \leftarrow \textsf{correction}\times \mathtt{flip}(\textsf{qubits})$
\ENDFOR

\STATE
\RETURN $\textsf{correction}$

\end{algorithmic}
\end{algorithm}

The two decoders that decode lineon and planeon defects differ only by the $\mathtt{neutralize}$ function, as well as the symmetries $\Gamma$ of the stabilizer group used to measure defects that have already been defined in the main text. In Algorithms~\ref{Alg:Lineon} and~\ref{Alg:Planeon} we describe the neutralize functions for the decoder for the lineon and planeon decoders, respectively. We assume that a cluster passed to either neutralize function has only topologically trivial paths with respect to the manifold on which the code is embedded. We also assume that the cluster is small with respect to the size of the system. We specify a subroutine $\mathtt{move}$ that returns a Pauli operator that moves a defect or defects from a location or set of locations to another specified location or set of locations along the shortest possible path. Calling the move function also updates the defect configuration of the cluster according to the move. For convenience we used colored superscripts to indicate the color of the string-like Pauli operator used to move the defects in Algorithm~\ref{Alg:Lineon}.

In Algorithm~\ref{Alg:Lineon} we specify planes $M^\mathbf{r}$, $M^\mathbf{g}$ and $M^\mathbf{b}$ that lie close to the cluster that are oriented orthogonal to the $y$ $z$ and $x$ axes respectively. With a minor abuse of notation we suppose that manifold $M^{\mathbf{r}}$ lies at constant $y = M^{\mathbf{r}} $, that manifold $M^{\mathbf{g}}$ lies at $z = M^{\mathbf{g}} $ and that $M^{\mathbf{b}}$ lies at $x = M^{\mathbf{b}} $, where the manifolds also represent constant numbers. The three parts of Algorithm~\ref{Alg:Lineon} correspond to Figs.~\ref{Fig:LineonCorrection}~(b),~(c) and~(d), respectively.

Similarly for Algorithm~\ref{Alg:Planeon} we specify a plane $M$ that lies at constant $z = M$ close to the input cluster. it will be helpful to subdivide the edges of an input cluster into subsets $\textsf{edges}^x$,$\textsf{edges}^y$ and $\textsf{edges}^z$ where the superscript $x$, $y$, $z$ indicates the orientation of the plane where the minimum-weight perfect matching subroutine was used to obtain the edge. The function of Algorithm~\ref{Alg:Planeon} is summarised in Fig.~\ref{Fig:PlaneonCorrection}.

\section{Relations, materialized symmetries and generalized Gauss's law}
\label{App:Relations}

We consider stabilizer codes that are specified by a set of generators that are local with respect to some natural lattice metric. For example consider a finite number of qubits on each site of a cubic lattice, and generators supported on cubes. 

\begin{defn}[Relation]
Any nontrivial product of stabilizer generators that equals the identity is called a relation. 
\end{defn}

\begin{defn}[Materialized symmetry]
Each relation implies a materialized symmetry of parity conservation for the set of all defects that excite an odd number of stabilizer generators that are involved in the relation.
\end{defn}
Materialized symmetries should not be confused with physical symmetries, which are operators that commute with the stabilizer generators and hence the Hamiltonian. However, it should be mentioned that physical symmetries lead to materialized symmetries when they are gauged~\cite{wegner1971duality, Vijay16, PhysRevB, kubica2018ungauging, you2018symmetric, shirley2018FoliatedFracton}. In particular, it was shown in Ref.~\cite{Kitaev03} that materialized symmetries are a consequence of the gauge symmetries that occur when a physical symmetry is gauged.  Throughout this paper we have exclusively discussed materialized symmetries, sometimes referring to them just as symmetries. 

\begin{defn}[Local relation]
For any subset of stabilizer generators whose elements have support contained inside a ball $R$, nontrivial relations among these stabilizers are called local relations.
\end{defn}
Here, $R$ is only required to be topologically equivalent to a ball, and also to be of constant extent as the lattice size is scaled up, or finite extent in an infinite lattice. 
 If we instead consider the larger subset of stabilizer generators that have support intersecting a ball $R$, we can define the group generated by products of these generators that leave no support on $R$. This clearly contains the local relations, and since both groups are abelian we can mod out by the local relations. The quotient group is a set of generalized Gauss's laws that relate the excitations within a ball to stabilizers that can be measured just outside its boundary.
\begin{defn}[Generalized Gauss's law]
A product of stabilizer generators, each with support intersecting a ball $R$, that has a nonempty support contained within the complement $R^c$ gives rise to a generalized Gauss's law on $R$.
\end{defn}
We remark that these Gauss's laws are only defined up to local relations. The above definition generalizes directly to any $R$ that is a connected proper subset. 

\begin{defn}[Global relation]
If the region $R$ is instead taken to be a closed manifold with no boundary, e.g. a sphere or torus, global relations are similarly defined to be nontrivial products of stabilizers that equal identity, modulo local relations. 
\end{defn}
Every global relation, when restricted to a ball, yields a Gauss's law. But not every Gauss's law on a finite ball leads to a global relation, it depends upon the boundary conditions. 
\begin{exmp}[Toric code]
For the two-dimensional toric code~\cite{Kitaev03} there are two independent global relations. These are given by the product of all star terms, and the product of all plaquette terms, respectively. These imply a materialized symmetry for the $e$ and $m$ excitations, namely that the total number of each type of excitation is even. 
Similarly, there are two independent Gauss's laws on a disc, one given by the product of all star terms overlapping the disc and the other given by the product of all plaquette terms. These relate the parity of $e$ and $m$ excitations within the disc to the eigenvalue of $X$ and $Z$ string operators on its boundary. This generalizes directly to any abelian quantum double~\cite{haah}.  
\end{exmp}
In a model that fulfills the topological order condition TQO2 in Ref.~\cite{Haah2013}, a cluster of charges that satisfies all boundary stabilizers involved in the Gauss's law for the smallest ball containing it is a topologically trivial superselection sector and hence can be created via a local operator in some (slightly larger) ball containing the charges. 
In the 2D toric code and the X-cube model we have considered, all the local Gauss's laws lead to global relations, and hence matching using all the global relations is sufficient to identify if a charge cluster is neutral. 

More generally, not every local Gauss's law will lead to a global relation and hence matching via only global relations may be insufficient to identify neutral clusters. So while the basic philosophy of our decoding strategy (pairing to satisfy all Gauss's laws) applies to any topological stabilizer code, the decoder must be adapted to capture the Gauss's laws locally if it is to succeed in a more general setting. We plan to explore this further in future work. 
It would also be interesting to determine whether our approach can be further generalized to other low-density parity-check codes. 

\begin{exmp}[The planar code]
Decoding the planar code~\cite{Dennis02} or, more generally, any toric code model with boundaries using minimum-weight perfect matching can be understood in terms of its symmetries. For such a model, where single defects can be created close to a boundary, the charge parity conservation law is not strictly respected. Instead, it is only respected up to the number of charges absorbed by the boundary.

We recover a materialized symmetry by over-completing the stabilizer group. For a system with boundaries, the product of all the star operators gives an operator that measures the parity of charges that have been absorbed by all the rough boundaries of the system. The inclusion of this operator, together with the star operators of the code, restore the symmetry of the system. 

In practice, to use minimum-weight perfect matching to decode this model, we first count all the defects on the system and, if we find an odd number of defects, we add an additional defect that lies `beyond' the boundary of the lattice to restore the even charge parity. We then permit defects close to the boundary to pair with this external charge. By counting all of the charges of the system, we are effectively inferring the total charge absorbed by the boundary. 

This example shows us that it is important to over complete the generating set of the stabilizer group to find materialized symmetries. Indeed, we use an over complete set of stabilizer generators to find the symmetries of the toric code on a lattice without boundaries. It is an interesting question to consider how the different ways of over completing the stabilizer group help us to find different decoding strategies.

\end{exmp}

\begin{exmp}[Wen's plaquette model]
To decode Wen's plaquette model~\cite{Wen2003}  one seeks to separately pair excitations on the black and white squares of a checkerboard lattice, however this is not globally well defined if the model is placed on a cylinder of odd circumference. 
\end{exmp}
\begin{exmp}[The color code]
The color code has a number of materialized symmetries. In essence, for all the variations of minimum-weight perfect matching decoders for the color code, the bit-flip error model is commonly decoded using matching subroutines that pair defects of two of its three differently colored types~\cite{Fowler11, Bombin12b, Delfosse14, Kubica18}, as these matchings respect a symmetry of the model. The union of the edges from two matching subroutines over the defects of two differently colored pairs of defect types gives neutralizable clusters. Lesser known are the fermionic materialized symmetries of the color code, see Appendix B in Ref.~\cite{Kesselring18}. Indeed the product of its Pauli-X stabililizers on the red plaquettes, Pauli-Y type stabilizers on the green plaquettes and Pauli-Z stabilizers on the blue plaquettes also returns identity. Permutations thereupon can be combined to find neutralizable clusters. One might consider the use of these symmetries to find a color code decoder subject to depolarizing noise. Matching decoders for higher-dimensional generalizations of the color code also rely on symmetries of the model~\cite{Aloshious18, Kubica18}.
\end{exmp}
\begin{exmp}[Restricted error models]
For some codes one can make use of additional effective symmetries when a restricted error model is considered. For instance the standard surface code that only experiences Pauli-Y errors effectively has one-dimensional materialized symmetries. The symmetries can be extended to two dimensional space-time planes when measurement errors are also modeled. This observation has lead to the design of a parallelized decoder to decode highly-biased errors using matching~\cite{Tuckett19}. Further, certain correlated error models can also be regarded as symmetry respecting. The ballistic error model proposed in Ref.~\cite{Nickerson17} respects an effective symmetry that enables the model to be decoded by mapping the code onto several decoupled copies of the surface code. 
\end{exmp}
\begin{exmp}[Two-dimensional Ising model]
The square lattice Ising model, in two dimensions, is a classical self-correcting memory due to the linear scaling of the energy penalty of an error region with the size of its boundary. It can also be decoded by a local cellular automata. This can be understood in terms of the extensive number of local relations in the model, one for each plaquette. 
Applying our approach to decode this model would lead to a local, parallelized decoder, capable of identifying neutral clusters. 
The same intuition extends to find a local decoder for the loop excitations of the three-dimensional toric code, and the four-dimensional toric code with loop excitations. 
We remark that local decoders for self correcting memories have also been found using machine-learning techniques~\cite{Breuckmann18}.
\end{exmp}

\begin{exmp}[Unreliable stabilizer measurements]
\label{Ex:measurementErrors}
In a practical setting, we are interested in the situation where a stabilizer measurement $S$ returns the incorrect outcome due to the fact that the measurement apparatus malfunctions. With this error model, we can obtain any syndrome configuration after measuring all the stabilizers of the system and, as such, we should not expect the syndrome to respect the materialized symmetries of the stabilizer group. To deal with this, we find that we can recover a materialized symmetry by extending the system along the time direction. This strategy was proposed in the seminal work of Dennis {\it et al.}~\cite{Dennis02} for the toric code. It was then tested for the toric code in Ref.~\cite{Wang03}. In fact, we expect this strategy holds generically for any decoder where neutral clusters of defects are found by matching pairs of defects according to their symmetries.

\begin{figure}
\includegraphics{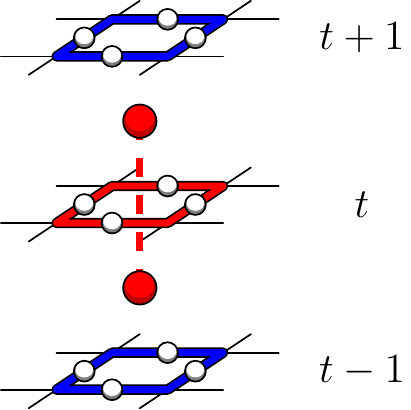}
\caption{A single stabilizer $S(T)$ that is measured sequentially at times $T = t-1,\,t,\,t+1$, where time is represented along the vertical direction. A measurement error occurs at time $t$. This creates defects for measurements $\tilde{S}(t-1) = S(t-1)S(t)$ and $\tilde{S}(t) = S(t) S(t+1)$. These defects are marked by red points. The measurement error can be viewed as a string segment that connects these defects along the temporal direction. \label{Fig:NoisyMeasurement}}
\end{figure}

To identify measurement errors, stabilizers measurements $S$ are repeated over time. It will be helpful to give a new presentation of the syndrome data that is collected over time. We denote the outcome of a measurement of stabilizer $S$ at time $t$ by $S(t) = \pm 1$. Assuming no errors occur, we have that $S(t)=+1$ for all $S$ and $t$. We obtain a syndrome by finding the parity of outcomes of pairs of time adjacent measurements. We thus define $\tilde{S}(t) = S(t)S(t+1)$ for all $S$ and $ t $, and we say that we find a defect for stabilizer $S$ at time $t$ if $\tilde{S}(t) = -1$.

Presenting the syndrome this way automatically extends the materialized symmetries of the model along the time direction, thus enabling to adapt the methods of decoding we have discussed in the main text. This is illustrated in Fig.~\ref{Fig:NoisyMeasurement}. Indeed, suppose a measurement error occurs such that we, incorrectly, observe $S(t) = -1$, and $S(T) = +1$ for all $T \not= t$. This will create two parity violations where we have that both $\tilde{S}(t-1) = -1$ and $\tilde{S}(t)= -1$. As precisely two defects are created with any given measurement error, we observe a defect parity conservation symmetry whereby measurement errors appear as strings that extend along the time direction where defects are created at their endpoints. As such, it is natural  to identify a likely configuration of measurement errors with a matching algorithm using this syndrome.

 Likewise, suppose a Pauli error $E$ occurs on a code state at time shortly after $t$ such that, assuming ideal measurements over time, we have that $S(T) = -1$ for all $T > t$ for all stabilizers satisfying $SE = -ES$. We have that $S(t)S(t+1) = -1$ for this subset of stabilizers and $S(T)S(T+1) = +1$ for all $T\not= t$. Clearly, we can use the symmetry-based decoding method to decode this error, as the defects created by $E$ must respect the materialized symmetries of the stabilizer group. We can therefore use matching to find neutral collections of defects by pairing according to the materialized symmetries of code. Indeed, decoding the defects on this constant time plane is equivalent to decoding in the ideal setting where all stabilizer measurements return the correct outcomes.

We require symmetries in this spacetime model to find neutralizable clusters of defects. In general, given a symmetry $\Sigma$ in the ideal case, whereby some subset of stabilizers $S \in \Sigma \subseteq \mathcal{S}$ satisfies $\prod_{S\in\Sigma} S = 1$, then we also have that $ \prod_{t,\, S\in \Sigma} \tilde{S}(t) = 1$. It is therefore possible to match defects according to these symmetries using the defects that are stretched out over spacetime. As an example, in Fig.~\ref{Fig:SpaceTimeSyndrome} we show a syndrome for the toric code in the presence of measurement noise. Measurement errors run as strings along the vertical direction of the figure, and physical errors lie as strings along planes that are orthogonal to the temporal direction. Defects are always created in pairs at the end points of the strings, thus respecting the symmetry.

\begin{figure}
\includegraphics{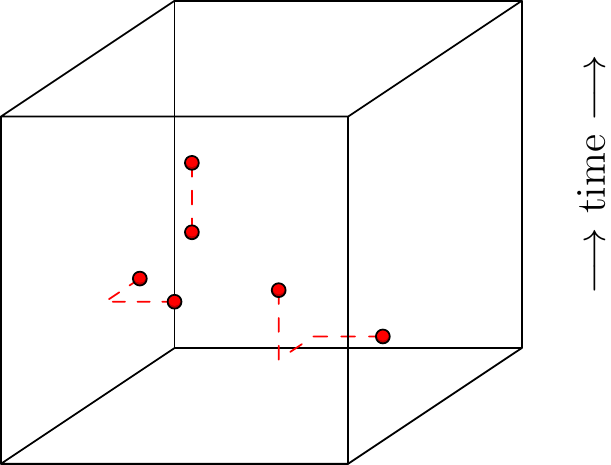}
\caption{A syndrome configuration for the toric code in the case where measurement errors are noisy. Measurement errors appear as strings that align with the temporal direction of the spacetime diagram, where time is running vertically along the page, and physical errors are strings that align along planes that are orthogonal to the temporal direction. Defects appear at the end points of the strings. \label{Fig:SpaceTimeSyndrome}}
\end{figure}

Given the symmetries that we have proposed, we must also find a correction that will reverse the error that gave rise to the syndrome. We assume that, in the ideal case, we know that matching on some set of symmetries $\Sigma$ will group defects into neutralizable clusters and, moreover, that we can find a correction for any such neutralizable cluster. Given these assumptions, it is easy to find a correction in the case where measurements are unreliable. Given a set of defects that are matched according to the spacetime symmetries, we first move all the defects of a neutral cluster onto a common constant-time plane that is near to all the defects, assuming that some series of measurement errors moved these defects away from this plane. With all the defects on any such plane we can apply a correction to all the qubits on this plane, as though no measurement errors had occured, as in the ideal case.

The argument for why this approach must work is as follows. A physical error will create a set of defects that respect the symmetries of the system, where all the defects lie at a common time plane. Measurement errors simply move these defects onto other time planes. We can determine the error that has occurred easily by moving all of these defects back onto a common plane. One may worry that problems may arise if we estimate the time an error occurred incorrectly. However, one need not worry. We can account for the effect of the error on the physical qubits provided we only know at what time the error occurred approximately. In general, one should not expect that they can determine the precise instant that an error has occurred. One should only aim for the more modest goal of correcting for errors on qubits within some sensible temporal proximity of the time the error occurred.  

It is worth adding that this method of decoding generalizes easily to foliated models in the sense of Refs.~\cite{Bolt16, Brown18}, the seminal example of which is the topological cluster state model due to Raussendorf {\it et al.}~\cite{Raussendorf05}. These methods were also used in a nontrivial setting recently~\cite{Tuckett19}. In a sense, the example presented in this reference for infinite noise bias is a classical two-dimensional variant of a fracton model, similar to that which we have studied here.

\end{exmp}

\section{Decoding time-correlated errors using parallelized quantum error correction}

\label{App:TimeCorrelated}

\begin{figure}
\includegraphics{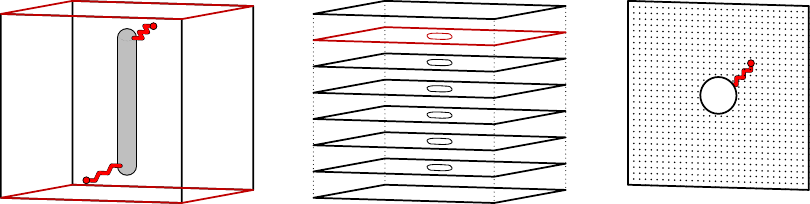}
\caption{(left) A spacetime diagram of a two-dimensional topological code, for instance the toric code, where time runs up along the page. We imagine an error that persists over a long time that will effectively introduce a puncture that extends through the time dimension. One can imagine a situation where a small error will have a catastrophic effect on the logical information. We show two small errors that introduce a defect into the extended puncture at the top and the bottom of the figure. These defects could easily be mismatched. (middle) We imagine a code where decoding can be parallelized on planes of constant time, i.e., a single-shot code. Again time runs up the page. In this situation the puncture that extends through time appears as a small puncture on each plane. Such a puncture is not disastrous to the decoding procedure. (right) We show the decoding problem on a single plane. The small puncture does not extend across the breadth of the plane so it is easy for a decoder to determine if a defect is absorbed by the small puncture. \label{Fig:TimeCorrelatedErrors}}
\end{figure}

In Ref.~\cite{Bombin16} it  was proven that a decoder that performs single-shot error correction~\cite{Bombin15a} can tolerate time correlated errors. This could be due to some malfunctioning hardware that requires maintenance. Here we give a simplified explanation of this result from the perspective of parallelized quantum error correction. We suggest that the threshold theorem against time correlated errors can be generalized to other types of correlated errors provided a code is chosen appropriately such that it can be decoded along parallelized planes that run orthogonal to the direction of the correlation. We also add that we can find solutions to this issue for codes that are not parallelizable. However, this needs to be conducted at the hardware level~\cite{Stace09, Stace10, Barrett10}, i.e. the circuit used to measure stabilizer data needs to be adapted such that stabilizer data is collected from gauge degrees of freedom or `junk qubits', see also more recent works~\cite{Auger17, Nagayama17}. Parallelized quantum error correction permits a solution at the software level.

The left of Fig.~\ref{Fig:TimeCorrelatedErrors} shows the spacetime history of the syndrome data of a code that does not readily permit parallelization, for instance the toric code. Suppose some components of the lattice are defective such that they experience errors at a very high frequency, or perhaps a stabilizer cannot be measured over many error-correction cycles. Such an error could be regarded as a puncture that extends a long distance over the time direction of spacetime. An extensive puncture will leave the encoded information vulnerable to low-weight errors if we do not take additional measures to identify them. We show a low-weight error in red where defects are separated over a long distance by this puncture. This could easily be mismatched by this decoder. It is important to identify the charge absorbed by the puncture. In the case of the toric code this can be solved by measuring superoperators to identify errors~\cite{Stace09, Stace10, Barrett10, Auger17, Nagayama17}. This may require altering the stabilizer measurement schedule. Temporally parallelized decoders can deal with such an error at the software level where we simply do not measure the stabilizers local to this time-correlated error.

The middle image in Fig.~\ref{Fig:TimeCorrelatedErrors} depicts the planes of constant time upon which the decoding subroutines of a single-shot decoder will act. Notably, the puncture is separated across all of these planes such that the puncture is very small in each subroutine. A small puncture can be dealt with easily on each plane. The right of Fig.~\ref{Fig:TimeCorrelatedErrors} shows how a small error close to the puncture will appear on a single constant time plane. A decoder will simply match the red defect to the small puncture in the centre of the lattice.

It may be interesting to extend the result due to Bomb\'{i}n to models with more symmetry, or symmetries other than the materialized symmetries of single-shot codes. One might expect that it is possible to show that self-correcting stabilizer codes~\cite{Alicki10, Brown16} demonstrate a threshold against error models that are correlated in an arbitrary spatial or temporal direction. We also note of recent work where a classical fractal code is demonstrated to be robust to one-dimensional errors using numerical experiments~\cite{Nixon20}.

\bibstyle{plain}


\begin{thebibliography}{72}
\expandafter\ifx\csname natexlab\endcsname\relax\def\natexlab#1{#1}\fi
\expandafter\ifx\csname bibnamefont\endcsname\relax
  \def\bibnamefont#1{#1}\fi
\expandafter\ifx\csname bibfnamefont\endcsname\relax
  \def\bibfnamefont#1{#1}\fi
\expandafter\ifx\csname citenamefont\endcsname\relax
  \def\citenamefont#1{#1}\fi
\expandafter\ifx\csname url\endcsname\relax
  \def\url#1{\texttt{#1}}\fi
\expandafter\ifx\csname urlprefix\endcsname\relax\def\urlprefix{URL }\fi
\providecommand{\bibinfo}[2]{#2}
\providecommand{\eprint}[2][]{\url{#2}}

\bibitem[{\citenamefont{Terhal}(2015)}]{Terhal15}
\bibinfo{author}{\bibfnamefont{B.~M.} \bibnamefont{Terhal}},
  \bibinfo{journal}{Rev. Mod. Phys.} \textbf{\bibinfo{volume}{87}},
  \bibinfo{pages}{307} (\bibinfo{year}{2015}).

\bibitem[{\citenamefont{Campbell et~al.}(2017)\citenamefont{Campbell, Terhal,
  and Vuillot}}]{Campbell17}
\bibinfo{author}{\bibfnamefont{E.~T.} \bibnamefont{Campbell}},
  \bibinfo{author}{\bibfnamefont{B.~M.} \bibnamefont{Terhal}},
  \bibnamefont{and} \bibinfo{author}{\bibfnamefont{C.}~\bibnamefont{Vuillot}},
  \bibinfo{journal}{Nature} \textbf{\bibinfo{volume}{549}},
  \bibinfo{pages}{172} (\bibinfo{year}{2017}).

\bibitem[{\citenamefont{Kitaev}(2003)}]{Kitaev03}
\bibinfo{author}{\bibfnamefont{A.~Y.} \bibnamefont{Kitaev}},
  \bibinfo{journal}{Ann. Phys.} \textbf{\bibinfo{volume}{303}},
  \bibinfo{pages}{2} (\bibinfo{year}{2003}).

\bibitem[{\citenamefont{Brown et~al.}(2016{\natexlab{a}})\citenamefont{Brown,
  Loss, Pachos, Self, and Wootton}}]{Brown16}
\bibinfo{author}{\bibfnamefont{B.~J.} \bibnamefont{Brown}},
  \bibinfo{author}{\bibfnamefont{D.}~\bibnamefont{Loss}},
  \bibinfo{author}{\bibfnamefont{J.~K.} \bibnamefont{Pachos}},
  \bibinfo{author}{\bibfnamefont{C.~N.} \bibnamefont{Self}}, \bibnamefont{and}
  \bibinfo{author}{\bibfnamefont{J.~R.} \bibnamefont{Wootton}},
  \bibinfo{journal}{Rev. Mod. Phys.} \textbf{\bibinfo{volume}{88}},
  \bibinfo{pages}{045005} (\bibinfo{year}{2016}{\natexlab{a}}).

\bibitem[{\citenamefont{Vijay et~al.}(2015)\citenamefont{Vijay, Haah, and
  Fu}}]{Vijay15}
\bibinfo{author}{\bibfnamefont{S.}~\bibnamefont{Vijay}},
  \bibinfo{author}{\bibfnamefont{J.}~\bibnamefont{Haah}}, \bibnamefont{and}
  \bibinfo{author}{\bibfnamefont{L.}~\bibnamefont{Fu}}, \bibinfo{journal}{Phys.
  Rev. B} \textbf{\bibinfo{volume}{92}}, \bibinfo{pages}{235136}
  (\bibinfo{year}{2015}).

\bibitem[{\citenamefont{Vijay et~al.}(2016)\citenamefont{Vijay, Haah, and
  Fu}}]{Vijay16}
\bibinfo{author}{\bibfnamefont{S.}~\bibnamefont{Vijay}},
  \bibinfo{author}{\bibfnamefont{J.}~\bibnamefont{Haah}}, \bibnamefont{and}
  \bibinfo{author}{\bibfnamefont{L.}~\bibnamefont{Fu}}, \bibinfo{journal}{Phys.
  Rev. B} \textbf{\bibinfo{volume}{94}}, \bibinfo{pages}{235157}
  (\bibinfo{year}{2016}).

\bibitem[{\citenamefont{Chamon}(2005)}]{Chamon05}
\bibinfo{author}{\bibfnamefont{C.}~\bibnamefont{Chamon}},
  \bibinfo{journal}{Phys. Rev. Lett.} \textbf{\bibinfo{volume}{94}},
  \bibinfo{pages}{040402} (\bibinfo{year}{2005}).

\bibitem[{\citenamefont{Haah}(2011)}]{Haah11}
\bibinfo{author}{\bibfnamefont{J.}~\bibnamefont{Haah}}, \bibinfo{journal}{Phys.
  Rev. A} \textbf{\bibinfo{volume}{83}}, \bibinfo{pages}{042330}
  (\bibinfo{year}{2011}).

\bibitem[{\citenamefont{Castelnovo and Chamon}(2011)}]{Castelnovo11}
\bibinfo{author}{\bibfnamefont{C.}~\bibnamefont{Castelnovo}} \bibnamefont{and}
  \bibinfo{author}{\bibfnamefont{C.}~\bibnamefont{Chamon}},
  \bibinfo{journal}{Phil. Mag.} \textbf{\bibinfo{volume}{92}},
  \bibinfo{pages}{1} (\bibinfo{year}{2011}).

\bibitem[{\citenamefont{Bravyi and Haah}(2011)}]{Bravyi11b}
\bibinfo{author}{\bibfnamefont{S.}~\bibnamefont{Bravyi}} \bibnamefont{and}
  \bibinfo{author}{\bibfnamefont{J.}~\bibnamefont{Haah}},
  \bibinfo{journal}{Phys. Rev. Lett.} \textbf{\bibinfo{volume}{107}},
  \bibinfo{pages}{150504} (\bibinfo{year}{2011}).

\bibitem[{\citenamefont{Bravyi and Haah}(2013)}]{Bravyi13a}
\bibinfo{author}{\bibfnamefont{S.}~\bibnamefont{Bravyi}} \bibnamefont{and}
  \bibinfo{author}{\bibfnamefont{J.}~\bibnamefont{Haah}},
  \bibinfo{journal}{Phys. Rev. Lett.} \textbf{\bibinfo{volume}{111}},
  \bibinfo{pages}{200501} (\bibinfo{year}{2013}).

\bibitem[{\citenamefont{Prem et~al.}(2017)\citenamefont{Prem, Haah, and
  Nandkishore}}]{PhysRevB.95.155133}
\bibinfo{author}{\bibfnamefont{A.}~\bibnamefont{Prem}},
  \bibinfo{author}{\bibfnamefont{J.}~\bibnamefont{Haah}}, \bibnamefont{and}
  \bibinfo{author}{\bibfnamefont{R.}~\bibnamefont{Nandkishore}},
  \bibinfo{journal}{Physical Review B} \textbf{\bibinfo{volume}{95}},
  \bibinfo{pages}{155133} (\bibinfo{year}{2017}), \eprint{1702.02952}.

\bibitem[{\citenamefont{Roberts and Bartlett}(2018)}]{Roberts18}
\bibinfo{author}{\bibfnamefont{S.}~\bibnamefont{Roberts}} \bibnamefont{and}
  \bibinfo{author}{\bibfnamefont{S.~D.} \bibnamefont{Bartlett}},
  \bibinfo{journal}{arXiv:1805.01474}  (\bibinfo{year}{2018}).

\bibitem[{\citenamefont{Schmitz et~al.}(2018)\citenamefont{Schmitz, Ma,
  Nandkishore, and Parameswaran}}]{PhysRevB.97.134426}
\bibinfo{author}{\bibfnamefont{A.~T.} \bibnamefont{Schmitz}},
  \bibinfo{author}{\bibfnamefont{H.}~\bibnamefont{Ma}},
  \bibinfo{author}{\bibfnamefont{R.~M.} \bibnamefont{Nandkishore}},
  \bibnamefont{and} \bibinfo{author}{\bibfnamefont{S.~A.}
  \bibnamefont{Parameswaran}}, \bibinfo{journal}{Physical Review B}
  \textbf{\bibinfo{volume}{97}}, \bibinfo{pages}{134426}
  (\bibinfo{year}{2018}), \eprint{1712.02375}.

\bibitem[{\citenamefont{Wegner}(1971)}]{wegner1971duality}
\bibinfo{author}{\bibfnamefont{F.~J.} \bibnamefont{Wegner}},
  \bibinfo{journal}{Journal of Mathematical Physics}
  \textbf{\bibinfo{volume}{12}}, \bibinfo{pages}{2259} (\bibinfo{year}{1971}).

\bibitem[{\citenamefont{Williamson}(2016)}]{PhysRevB}
\bibinfo{author}{\bibfnamefont{D.~J.} \bibnamefont{Williamson}},
  \bibinfo{journal}{Physical Review B} \textbf{\bibinfo{volume}{94}},
  \bibinfo{pages}{155128} (\bibinfo{year}{2016}), \eprint{1603.05182}.

\bibitem[{\citenamefont{Kubica and Yoshida}(2018)}]{kubica2018ungauging}
\bibinfo{author}{\bibfnamefont{A.}~\bibnamefont{Kubica}} \bibnamefont{and}
  \bibinfo{author}{\bibfnamefont{B.}~\bibnamefont{Yoshida}}, pp.
  \bibinfo{pages}{1--33} (\bibinfo{year}{2018}), \eprint{arXiv:1805.01836v1}.

\bibitem[{\citenamefont{You et~al.}(2018)\citenamefont{You, Devakul, Burnell,
  and Sondhi}}]{you2018symmetric}
\bibinfo{author}{\bibfnamefont{Y.}~\bibnamefont{You}},
  \bibinfo{author}{\bibfnamefont{T.}~\bibnamefont{Devakul}},
  \bibinfo{author}{\bibfnamefont{F.~J.} \bibnamefont{Burnell}},
  \bibnamefont{and} \bibinfo{author}{\bibfnamefont{S.~L.} \bibnamefont{Sondhi}}
  (\bibinfo{year}{2018}), \eprint{1805.09800}.

\bibitem[{\citenamefont{Shirley
  et~al.}(2018{\natexlab{a}})\citenamefont{Shirley, Slagle, and
  Chen}}]{shirley2018FoliatedFracton}
\bibinfo{author}{\bibfnamefont{W.}~\bibnamefont{Shirley}},
  \bibinfo{author}{\bibfnamefont{K.}~\bibnamefont{Slagle}}, \bibnamefont{and}
  \bibinfo{author}{\bibfnamefont{X.}~\bibnamefont{Chen}}
  (\bibinfo{year}{2018}{\natexlab{a}}), \eprint{arXiv:1806.08679v1}.

\bibitem[{\citenamefont{Castelnovo et~al.}(2010)\citenamefont{Castelnovo,
  Chamon, and Sherrington}}]{PhysRevB.81.184303}
\bibinfo{author}{\bibfnamefont{C.}~\bibnamefont{Castelnovo}},
  \bibinfo{author}{\bibfnamefont{C.}~\bibnamefont{Chamon}}, \bibnamefont{and}
  \bibinfo{author}{\bibfnamefont{D.}~\bibnamefont{Sherrington}},
  \bibinfo{journal}{Phys. Rev. B - Condens. Matter Mater. Phys.}
  \textbf{\bibinfo{volume}{81}}, \bibinfo{pages}{184303}
  (\bibinfo{year}{2010}), \eprint{1003.3832}.

\bibitem[{\citenamefont{Gottesman}(2001)}]{GottesmanThesis}
\bibinfo{author}{\bibfnamefont{D.}~\bibnamefont{Gottesman}}, Ph.D. thesis,
  \bibinfo{school}{California Institute of Technology} (\bibinfo{year}{2001}).

\bibitem[{\citenamefont{Bombin and Martin-Delgado}(2007)}]{Bombin07a}
\bibinfo{author}{\bibfnamefont{H.}~\bibnamefont{Bombin}} \bibnamefont{and}
  \bibinfo{author}{\bibfnamefont{M.~A.} \bibnamefont{Martin-Delgado}},
  \bibinfo{journal}{Phys. Rev. Lett.} \textbf{\bibinfo{volume}{98}},
  \bibinfo{pages}{160502} (\bibinfo{year}{2007}).

\bibitem[{\citenamefont{Kim}(2011)}]{Kim11}
\bibinfo{author}{\bibfnamefont{I.~H.} \bibnamefont{Kim}},
  \bibinfo{journal}{Phys. Rev. A} \textbf{\bibinfo{volume}{83}},
  \bibinfo{pages}{052308} (\bibinfo{year}{2011}).

\bibitem[{\citenamefont{Brown et~al.}(2016{\natexlab{b}})\citenamefont{Brown,
  Nickerson, and Browne}}]{Brown16a}
\bibinfo{author}{\bibfnamefont{B.~J.} \bibnamefont{Brown}},
  \bibinfo{author}{\bibfnamefont{N.~H.} \bibnamefont{Nickerson}},
  \bibnamefont{and} \bibinfo{author}{\bibfnamefont{D.~E.}
  \bibnamefont{Browne}}, \bibinfo{journal}{Nat. Commun.}
  \textbf{\bibinfo{volume}{7}}, \bibinfo{pages}{12302}
  (\bibinfo{year}{2016}{\natexlab{b}}).

\bibitem[{\citenamefont{Bombin and Martin-Delagado}(2006)}]{Bombin06}
\bibinfo{author}{\bibfnamefont{H.}~\bibnamefont{Bombin}} \bibnamefont{and}
  \bibinfo{author}{\bibfnamefont{M.~A.} \bibnamefont{Martin-Delagado}},
  \bibinfo{journal}{Phys. Rev. Lett.} \textbf{\bibinfo{volume}{97}},
  \bibinfo{pages}{180501} (\bibinfo{year}{2006}).

\bibitem[{\citenamefont{Harrington}(2004)}]{Harrington}
\bibinfo{author}{\bibfnamefont{J.~W.} \bibnamefont{Harrington}}, Ph.D. thesis,
  \bibinfo{school}{California Institute of Technology} (\bibinfo{year}{2004}).

\bibitem[{\citenamefont{Duclos-Cianci and Poulin}(2010)}]{Duclos-Cianci10}
\bibinfo{author}{\bibfnamefont{G.}~\bibnamefont{Duclos-Cianci}}
  \bibnamefont{and} \bibinfo{author}{\bibfnamefont{D.}~\bibnamefont{Poulin}},
  \bibinfo{journal}{Phys. Rev. Lett.} \textbf{\bibinfo{volume}{104}},
  \bibinfo{pages}{050504} (\bibinfo{year}{2010}).

\bibitem[{\citenamefont{Wootton and Loss}(2012)}]{Wootton12}
\bibinfo{author}{\bibfnamefont{J.~R.} \bibnamefont{Wootton}} \bibnamefont{and}
  \bibinfo{author}{\bibfnamefont{D.}~\bibnamefont{Loss}},
  \bibinfo{journal}{Phys. Rev. Lett.} \textbf{\bibinfo{volume}{109}},
  \bibinfo{pages}{160503} (\bibinfo{year}{2012}).

\bibitem[{\citenamefont{Anwar et~al.}(2014)\citenamefont{Anwar, Brown,
  Campbell, and Browne}}]{Anwar14}
\bibinfo{author}{\bibfnamefont{H.}~\bibnamefont{Anwar}},
  \bibinfo{author}{\bibfnamefont{B.~J.} \bibnamefont{Brown}},
  \bibinfo{author}{\bibfnamefont{E.~T.} \bibnamefont{Campbell}},
  \bibnamefont{and} \bibinfo{author}{\bibfnamefont{D.~E.}
  \bibnamefont{Browne}}, \bibinfo{journal}{New J. Phys.}
  \textbf{\bibinfo{volume}{16}}, \bibinfo{pages}{063038}
  (\bibinfo{year}{2014}).

\bibitem[{\citenamefont{Torlai and Melko}(2017)}]{Torlai17}
\bibinfo{author}{\bibfnamefont{G.}~\bibnamefont{Torlai}} \bibnamefont{and}
  \bibinfo{author}{\bibfnamefont{R.~G.} \bibnamefont{Melko}},
  \bibinfo{journal}{Phys. Rev. Lett.} \textbf{\bibinfo{volume}{119}},
  \bibinfo{pages}{030501} (\bibinfo{year}{2017}).

\bibitem[{\citenamefont{Delfosse and Nickerson}(2017)}]{Delfosse17}
\bibinfo{author}{\bibfnamefont{N.}~\bibnamefont{Delfosse}} \bibnamefont{and}
  \bibinfo{author}{\bibfnamefont{N.~H.} \bibnamefont{Nickerson}},
  \bibinfo{journal}{arXiv:1709.06218}  (\bibinfo{year}{2017}).

\bibitem[{\citenamefont{Kolmogorov}(2009)}]{Kolmogorov09}
\bibinfo{author}{\bibfnamefont{V.}~\bibnamefont{Kolmogorov}},
  \bibinfo{journal}{Math. Prog. Comp.} \textbf{\bibinfo{volume}{1}},
  \bibinfo{pages}{43} (\bibinfo{year}{2009}).

\bibitem[{\citenamefont{Edmonds}(1965)}]{Edmonds65}
\bibinfo{author}{\bibfnamefont{J.}~\bibnamefont{Edmonds}},
  \bibinfo{journal}{Canad. J. Math.} \textbf{\bibinfo{volume}{17}},
  \bibinfo{pages}{449} (\bibinfo{year}{1965}).

\bibitem[{\citenamefont{Dennis et~al.}(2002)\citenamefont{Dennis, Kitaev,
  Landahl, and Preskill}}]{Dennis02}
\bibinfo{author}{\bibfnamefont{E.}~\bibnamefont{Dennis}},
  \bibinfo{author}{\bibfnamefont{A.}~\bibnamefont{Kitaev}},
  \bibinfo{author}{\bibfnamefont{A.}~\bibnamefont{Landahl}}, \bibnamefont{and}
  \bibinfo{author}{\bibfnamefont{J.}~\bibnamefont{Preskill}},
  \bibinfo{journal}{J. Math. Phys.} \textbf{\bibinfo{volume}{43}},
  \bibinfo{pages}{4452} (\bibinfo{year}{2002}).

\bibitem[{\citenamefont{Stace and Barrett}(2010)}]{Stace10}
\bibinfo{author}{\bibfnamefont{T.~M.} \bibnamefont{Stace}} \bibnamefont{and}
  \bibinfo{author}{\bibfnamefont{S.~D.} \bibnamefont{Barrett}},
  \bibinfo{journal}{Phys. Rev. A} \textbf{\bibinfo{volume}{81}},
  \bibinfo{pages}{022317} (\bibinfo{year}{2010}).

\bibitem[{\citenamefont{Fowler et~al.}(2012)\citenamefont{Fowler, Whiteside,
  McInnes, and Rabbani}}]{Fowler12b}
\bibinfo{author}{\bibfnamefont{A.~G.} \bibnamefont{Fowler}},
  \bibinfo{author}{\bibfnamefont{A.~C.} \bibnamefont{Whiteside}},
  \bibinfo{author}{\bibfnamefont{A.~L.} \bibnamefont{McInnes}},
  \bibnamefont{and} \bibinfo{author}{\bibfnamefont{A.}~\bibnamefont{Rabbani}},
  \bibinfo{journal}{Phys. Rev. X} \textbf{\bibinfo{volume}{2}},
  \bibinfo{pages}{041003} (\bibinfo{year}{2012}).

\bibitem[{\citenamefont{Hutter and Loss}(2014)}]{Hutter14a}
\bibinfo{author}{\bibfnamefont{A.}~\bibnamefont{Hutter}} \bibnamefont{and}
  \bibinfo{author}{\bibfnamefont{D.}~\bibnamefont{Loss}},
  \bibinfo{journal}{Phys. Rev. A} \textbf{\bibinfo{volume}{89}},
  \bibinfo{pages}{042334} (\bibinfo{year}{2014}).

\bibitem[{\citenamefont{Nickerson and Brown}(2017)}]{Nickerson17}
\bibinfo{author}{\bibfnamefont{N.~H.} \bibnamefont{Nickerson}}
  \bibnamefont{and} \bibinfo{author}{\bibfnamefont{B.~J.} \bibnamefont{Brown}},
  \bibinfo{journal}{arXiv:1712.00502}  (\bibinfo{year}{2017}).

\bibitem[{\citenamefont{Criger and Ashraf}(2018)}]{Criger18}
\bibinfo{author}{\bibfnamefont{B.}~\bibnamefont{Criger}} \bibnamefont{and}
  \bibinfo{author}{\bibfnamefont{I.}~\bibnamefont{Ashraf}},
  \bibinfo{journal}{Quantum} \textbf{\bibinfo{volume}{2}}, \bibinfo{pages}{102}
  (\bibinfo{year}{2018}).

\bibitem[{\citenamefont{Beverland et~al.}(2018)\citenamefont{Beverland, Brown,
  Kastoryano, and Marolleau}}]{Beverland18}
\bibinfo{author}{\bibfnamefont{M.~E.} \bibnamefont{Beverland}},
  \bibinfo{author}{\bibfnamefont{B.~J.} \bibnamefont{Brown}},
  \bibinfo{author}{\bibfnamefont{M.~J.} \bibnamefont{Kastoryano}},
  \bibnamefont{and}
  \bibinfo{author}{\bibfnamefont{Q.}~\bibnamefont{Marolleau}},
  \bibinfo{journal}{arXiv:1812.05117}  (\bibinfo{year}{2018}).

\bibitem[{\citenamefont{Kubica et~al.}(2017)\citenamefont{Kubica, Beverland,
  Brandao, Preskill, and Svore}}]{Kubica17}
\bibinfo{author}{\bibfnamefont{A.}~\bibnamefont{Kubica}},
  \bibinfo{author}{\bibfnamefont{M.~E.} \bibnamefont{Beverland}},
  \bibinfo{author}{\bibfnamefont{F.}~\bibnamefont{Brandao}},
  \bibinfo{author}{\bibfnamefont{J.}~\bibnamefont{Preskill}}, \bibnamefont{and}
  \bibinfo{author}{\bibfnamefont{K.~M.} \bibnamefont{Svore}},
  \bibinfo{journal}{arXiv:1708.07131}  (\bibinfo{year}{2017}).

\bibitem[{\citenamefont{Bomb\'{i}n}(2015)}]{Bombin15a}
\bibinfo{author}{\bibfnamefont{H.}~\bibnamefont{Bomb\'{i}n}},
  \bibinfo{journal}{Phys. Rev. X} \textbf{\bibinfo{volume}{5}},
  \bibinfo{pages}{031043} (\bibinfo{year}{2015}).

\bibitem[{\citenamefont{Campbell}(2019)}]{Campbell18}
\bibinfo{author}{\bibfnamefont{E.~T.} \bibnamefont{Campbell}},
  \bibinfo{journal}{Quantum Science and Technology}
  \textbf{\bibinfo{volume}{4}}, \bibinfo{pages}{025006} (\bibinfo{year}{2019}),
  \eprint{arXiv:1805.09271}.

\bibitem[{\citenamefont{Fawzi et~al.}(2018)\citenamefont{Fawzi, Grospellier,
  and Leverrier}}]{fawzi2018constant}
\bibinfo{author}{\bibfnamefont{O.}~\bibnamefont{Fawzi}},
  \bibinfo{author}{\bibfnamefont{A.}~\bibnamefont{Grospellier}},
  \bibnamefont{and}
  \bibinfo{author}{\bibfnamefont{A.}~\bibnamefont{Leverrier}},
  \bibinfo{journal}{arXiv preprint arXiv:1808.03821}  (\bibinfo{year}{2018}).

\bibitem[{\citenamefont{Bomb\'{i}n}(2016)}]{Bombin16}
\bibinfo{author}{\bibfnamefont{H.}~\bibnamefont{Bomb\'{i}n}},
  \bibinfo{journal}{Phys. Rev. X} \textbf{\bibinfo{volume}{6}},
  \bibinfo{pages}{041034} (\bibinfo{year}{2016}).

\bibitem[{\citenamefont{Bolt et~al.}(2016)\citenamefont{Bolt, Duclos-Cianci,
  Poulin, and Stace}}]{Bolt16}
\bibinfo{author}{\bibfnamefont{A.}~\bibnamefont{Bolt}},
  \bibinfo{author}{\bibfnamefont{G.}~\bibnamefont{Duclos-Cianci}},
  \bibinfo{author}{\bibfnamefont{D.}~\bibnamefont{Poulin}}, \bibnamefont{and}
  \bibinfo{author}{\bibfnamefont{T.~M.} \bibnamefont{Stace}},
  \bibinfo{journal}{Phys. Rev. Lett.} \textbf{\bibinfo{volume}{117}},
  \bibinfo{pages}{070501} (\bibinfo{year}{2016}).

\bibitem[{\citenamefont{Brown and Roberts}(2018)}]{Brown18}
\bibinfo{author}{\bibfnamefont{B.~J.} \bibnamefont{Brown}} \bibnamefont{and}
  \bibinfo{author}{\bibfnamefont{S.}~\bibnamefont{Roberts}},
  \bibinfo{journal}{arXiv:1811.11780}  (\bibinfo{year}{2018}).

\bibitem[{\citenamefont{Pastawski et~al.}(2011)\citenamefont{Pastawski,
  Clemente, and Cirac}}]{Pastawski11}
\bibinfo{author}{\bibfnamefont{F.}~\bibnamefont{Pastawski}},
  \bibinfo{author}{\bibfnamefont{L.}~\bibnamefont{Clemente}}, \bibnamefont{and}
  \bibinfo{author}{\bibfnamefont{J.~I.} \bibnamefont{Cirac}},
  \bibinfo{journal}{Phys. Rev. A} \textbf{\bibinfo{volume}{83}},
  \bibinfo{pages}{012304} (\bibinfo{year}{2011}).

\bibitem[{\citenamefont{Breuckmann et~al.}(2017)\citenamefont{Breuckmann,
  Duivenvoorden, Michels, and Terhal}}]{Breuckmann17}
\bibinfo{author}{\bibfnamefont{N.~P.} \bibnamefont{Breuckmann}},
  \bibinfo{author}{\bibfnamefont{K.}~\bibnamefont{Duivenvoorden}},
  \bibinfo{author}{\bibfnamefont{D.}~\bibnamefont{Michels}}, \bibnamefont{and}
  \bibinfo{author}{\bibfnamefont{B.~M.} \bibnamefont{Terhal}},
  \bibinfo{journal}{Quant. Inf. Comp.} \textbf{\bibinfo{volume}{17}},
  \bibinfo{pages}{0181} (\bibinfo{year}{2017}).

\bibitem[{\citenamefont{Kubica and Preskill}(2018)}]{Kubica18}
\bibinfo{author}{\bibfnamefont{A.}~\bibnamefont{Kubica}} \bibnamefont{and}
  \bibinfo{author}{\bibfnamefont{J.}~\bibnamefont{Preskill}},
  \bibinfo{journal}{arXiv:1809.10145}  (\bibinfo{year}{2018}).

\bibitem[{\citenamefont{Shirley
  et~al.}(2018{\natexlab{b}})\citenamefont{Shirley, Slagle, Wang, and
  Chen}}]{shirley2017fracton}
\bibinfo{author}{\bibfnamefont{W.}~\bibnamefont{Shirley}},
  \bibinfo{author}{\bibfnamefont{K.}~\bibnamefont{Slagle}},
  \bibinfo{author}{\bibfnamefont{Z.}~\bibnamefont{Wang}}, \bibnamefont{and}
  \bibinfo{author}{\bibfnamefont{X.}~\bibnamefont{Chen}},
  \bibinfo{journal}{Physical Review X} \textbf{\bibinfo{volume}{8}}
  (\bibinfo{year}{2018}{\natexlab{b}}), \eprint{1712.05892}.

\bibitem[{\citenamefont{Shirley
  et~al.}(2018{\natexlab{c}})\citenamefont{Shirley, Slagle, and
  Chen}}]{shirley2018Fractional}
\bibinfo{author}{\bibfnamefont{W.}~\bibnamefont{Shirley}},
  \bibinfo{author}{\bibfnamefont{K.}~\bibnamefont{Slagle}}, \bibnamefont{and}
  \bibinfo{author}{\bibfnamefont{X.}~\bibnamefont{Chen}}
  (\bibinfo{year}{2018}{\natexlab{c}}), \eprint{1806.08625}.

\bibitem[{\citenamefont{Slagle et~al.}(2018)\citenamefont{Slagle, Aasen, and
  Williamson}}]{Slagle2018}
\bibinfo{author}{\bibfnamefont{K.}~\bibnamefont{Slagle}},
  \bibinfo{author}{\bibfnamefont{D.}~\bibnamefont{Aasen}}, \bibnamefont{and}
  \bibinfo{author}{\bibfnamefont{D.}~\bibnamefont{Williamson}}
  (\bibinfo{year}{2018}), \eprint{1812.01613}.

\bibitem[{\citenamefont{Nixon and Brown}(2020)}]{Nixon20}
\bibinfo{author}{\bibfnamefont{G.~M.} \bibnamefont{Nixon}} \bibnamefont{and}
  \bibinfo{author}{\bibfnamefont{B.~J.} \bibnamefont{Brown}},
  \bibinfo{journal}{arXiv:2002.11738}  (\bibinfo{year}{2020}).

\bibitem[{\citenamefont{Delfosse}(2014)}]{Delfosse14}
\bibinfo{author}{\bibfnamefont{N.}~\bibnamefont{Delfosse}},
  \bibinfo{journal}{Phys. Rev. A} \textbf{\bibinfo{volume}{89}},
  \bibinfo{pages}{012317} (\bibinfo{year}{2014}).

\bibitem[{\citenamefont{Kubica}(2018)}]{KubicaThesis}
\bibinfo{author}{\bibfnamefont{A.~M.} \bibnamefont{Kubica}}, Ph.D. thesis,
  \bibinfo{school}{California Institute of Technology} (\bibinfo{year}{2018}).

\bibitem[{\citenamefont{Wang et~al.}(2003)\citenamefont{Wang, Harrington, and
  Preskill}}]{Wang03}
\bibinfo{author}{\bibfnamefont{C.}~\bibnamefont{Wang}},
  \bibinfo{author}{\bibfnamefont{J.}~\bibnamefont{Harrington}},
  \bibnamefont{and} \bibinfo{author}{\bibfnamefont{J.}~\bibnamefont{Preskill}},
  \bibinfo{journal}{Ann. Phys.} \textbf{\bibinfo{volume}{303}},
  \bibinfo{pages}{31} (\bibinfo{year}{2003}).

\bibitem[{\citenamefont{Haah}(2016)}]{haah}
\bibinfo{author}{\bibfnamefont{J.}~\bibnamefont{Haah}},
  \bibinfo{journal}{Communications in Mathematical Physics}
  \textbf{\bibinfo{volume}{342}}, \bibinfo{pages}{771} (\bibinfo{year}{2016}),
  \eprint{1407.2926}.

\bibitem[{\citenamefont{Haah}(2013)}]{Haah2013}
\bibinfo{author}{\bibfnamefont{J.}~\bibnamefont{Haah}}, p. \bibinfo{pages}{200}
  (\bibinfo{year}{2013}), \eprint{1305.6973}.

\bibitem[{\citenamefont{Wen}(2003)}]{Wen2003}
\bibinfo{author}{\bibfnamefont{X.~G.} \bibnamefont{Wen}},
  \bibinfo{journal}{Physical Review Letters} \textbf{\bibinfo{volume}{90}},
  \bibinfo{pages}{4} (\bibinfo{year}{2003}), \eprint{0205004}.

\bibitem[{\citenamefont{Fowler}(2011)}]{Fowler11}
\bibinfo{author}{\bibfnamefont{A.~G.} \bibnamefont{Fowler}},
  \bibinfo{journal}{Phys. Rev. A} \textbf{\bibinfo{volume}{83}},
  \bibinfo{pages}{042310} (\bibinfo{year}{2011}).

\bibitem[{\citenamefont{Bombin et~al.}(2012)\citenamefont{Bombin,
  Duclos-Cianci, and Poulin}}]{Bombin12b}
\bibinfo{author}{\bibfnamefont{H.}~\bibnamefont{Bombin}},
  \bibinfo{author}{\bibfnamefont{G.}~\bibnamefont{Duclos-Cianci}},
  \bibnamefont{and} \bibinfo{author}{\bibfnamefont{D.}~\bibnamefont{Poulin}},
  \bibinfo{journal}{New J. Phys.} \textbf{\bibinfo{volume}{14}},
  \bibinfo{pages}{073048} (\bibinfo{year}{2012}).

\bibitem[{\citenamefont{Kesselring et~al.}(2018)\citenamefont{Kesselring,
  Pastawski, Eisert, and Brown}}]{Kesselring18}
\bibinfo{author}{\bibfnamefont{M.~S.} \bibnamefont{Kesselring}},
  \bibinfo{author}{\bibfnamefont{F.}~\bibnamefont{Pastawski}},
  \bibinfo{author}{\bibfnamefont{J.}~\bibnamefont{Eisert}}, \bibnamefont{and}
  \bibinfo{author}{\bibfnamefont{B.~J.} \bibnamefont{Brown}},
  \bibinfo{journal}{Quantum} \textbf{\bibinfo{volume}{2}}, \bibinfo{pages}{101}
  (\bibinfo{year}{2018}).

\bibitem[{\citenamefont{Aloshious and Sarvepalli}(2018)}]{Aloshious18}
\bibinfo{author}{\bibfnamefont{A.~B.} \bibnamefont{Aloshious}}
  \bibnamefont{and} \bibinfo{author}{\bibfnamefont{P.~K.}
  \bibnamefont{Sarvepalli}}, \bibinfo{journal}{Phys. Rev. A}
  \textbf{\bibinfo{volume}{98}}, \bibinfo{pages}{012302}
  (\bibinfo{year}{2018}).

\bibitem[{\citenamefont{Tuckett et~al.}(2020)\citenamefont{Tuckett, Bartlett,
  Flammia, and Brown}}]{Tuckett19}
\bibinfo{author}{\bibfnamefont{D.~K.} \bibnamefont{Tuckett}},
  \bibinfo{author}{\bibfnamefont{S.~D.} \bibnamefont{Bartlett}},
  \bibinfo{author}{\bibfnamefont{S.~T.} \bibnamefont{Flammia}},
  \bibnamefont{and} \bibinfo{author}{\bibfnamefont{B.~J.} \bibnamefont{Brown}},
  \bibinfo{journal}{Phys. Rev. Lett.} \textbf{\bibinfo{volume}{124}},
  \bibinfo{pages}{130501} (\bibinfo{year}{2020}).

\bibitem[{\citenamefont{Breuckmann and Ni}(2018)}]{Breuckmann18}
\bibinfo{author}{\bibfnamefont{N.~P.} \bibnamefont{Breuckmann}}
  \bibnamefont{and} \bibinfo{author}{\bibfnamefont{X.}~\bibnamefont{Ni}},
  \bibinfo{journal}{Quantum} \textbf{\bibinfo{volume}{2}}, \bibinfo{pages}{68}
  (\bibinfo{year}{2018}).

\bibitem[{\citenamefont{Raussendorf et~al.}(2005)\citenamefont{Raussendorf,
  Bravyi, and Harrington}}]{Raussendorf05}
\bibinfo{author}{\bibfnamefont{R.}~\bibnamefont{Raussendorf}},
  \bibinfo{author}{\bibfnamefont{S.}~\bibnamefont{Bravyi}}, \bibnamefont{and}
  \bibinfo{author}{\bibfnamefont{J.}~\bibnamefont{Harrington}},
  \bibinfo{journal}{Phys. Rev. A} \textbf{\bibinfo{volume}{71}},
  \bibinfo{pages}{062313} (\bibinfo{year}{2005}).

\bibitem[{\citenamefont{Stace et~al.}(2009)\citenamefont{Stace, Barrett, and
  Doherty}}]{Stace09}
\bibinfo{author}{\bibfnamefont{T.~M.} \bibnamefont{Stace}},
  \bibinfo{author}{\bibfnamefont{S.~D.} \bibnamefont{Barrett}},
  \bibnamefont{and} \bibinfo{author}{\bibfnamefont{A.~C.}
  \bibnamefont{Doherty}}, \bibinfo{journal}{Phys. Rev. Lett.}
  \textbf{\bibinfo{volume}{102}}, \bibinfo{pages}{200501}
  (\bibinfo{year}{2009}).

\bibitem[{\citenamefont{Barrett and Stace}(2010)}]{Barrett10}
\bibinfo{author}{\bibfnamefont{S.~D.} \bibnamefont{Barrett}} \bibnamefont{and}
  \bibinfo{author}{\bibfnamefont{T.~M.} \bibnamefont{Stace}},
  \bibinfo{journal}{Phys. Rev. Lett.} \textbf{\bibinfo{volume}{105}},
  \bibinfo{pages}{200502} (\bibinfo{year}{2010}).

\bibitem[{\citenamefont{Auger et~al.}(2017)\citenamefont{Auger, Anwar,
  Gimeno-Segovia, Stace, and Browne}}]{Auger17}
\bibinfo{author}{\bibfnamefont{J.~M.} \bibnamefont{Auger}},
  \bibinfo{author}{\bibfnamefont{H.}~\bibnamefont{Anwar}},
  \bibinfo{author}{\bibfnamefont{M.}~\bibnamefont{Gimeno-Segovia}},
  \bibinfo{author}{\bibfnamefont{T.~M.} \bibnamefont{Stace}}, \bibnamefont{and}
  \bibinfo{author}{\bibfnamefont{D.~E.} \bibnamefont{Browne}},
  \bibinfo{journal}{Phys. Rev. A} \textbf{\bibinfo{volume}{96}},
  \bibinfo{pages}{042316} (\bibinfo{year}{2017}).

\bibitem[{\citenamefont{Nagayama et~al.}(2017)\citenamefont{Nagayama, Fowler,
  Horsman, Devitt, and Meter}}]{Nagayama17}
\bibinfo{author}{\bibfnamefont{S.}~\bibnamefont{Nagayama}},
  \bibinfo{author}{\bibfnamefont{A.~G.} \bibnamefont{Fowler}},
  \bibinfo{author}{\bibfnamefont{D.}~\bibnamefont{Horsman}},
  \bibinfo{author}{\bibfnamefont{S.~J.} \bibnamefont{Devitt}},
  \bibnamefont{and} \bibinfo{author}{\bibfnamefont{R.~V.} \bibnamefont{Meter}},
  \bibinfo{journal}{New J. Phys.} \textbf{\bibinfo{volume}{19}},
  \bibinfo{pages}{023050} (\bibinfo{year}{2017}).

\bibitem[{\citenamefont{Alicki et~al.}(2010)\citenamefont{Alicki, Horodecki,
  Horodecki, and Horodecki}}]{Alicki10}
\bibinfo{author}{\bibfnamefont{R.}~\bibnamefont{Alicki}},
  \bibinfo{author}{\bibfnamefont{M.}~\bibnamefont{Horodecki}},
  \bibinfo{author}{\bibfnamefont{P.}~\bibnamefont{Horodecki}},
  \bibnamefont{and}
  \bibinfo{author}{\bibfnamefont{R.}~\bibnamefont{Horodecki}},
  \bibinfo{journal}{Open Syst. Inf. Dyn.} \textbf{\bibinfo{volume}{17}},
  \bibinfo{pages}{1} (\bibinfo{year}{2010}).

\end{thebibliography}

\end{document}